\documentclass[prd,amsmath,amssymb,floatfix,superscriptaddress,notitlepage,nofootinbib,preprintnumbers]{revtex4-1}
\usepackage{amsfonts,amssymb,amsmath,graphicx,color,bm,enumitem}
\definecolor{ultramarine}{rgb}{0.07, 0.04, 0.56}
\definecolor{cadmiumgreen}{rgb}{0.0, 0.42, 0.24}
\definecolor{indigo(dye)}{rgb}{0.0, 0.25, 0.42}
\usepackage[linktocpage=true]{hyperref}
\hypersetup{
colorlinks=true,
citecolor=red,
linkcolor=blue,
urlcolor=indigo(dye),
}

\newcommand{\be}{\begin{equation}}  
\newcommand{\ee}{\end{equation}}
\newcommand{\bem}{\begin{pmatrix}}
\newcommand{\eem}{\end{pmatrix}}

\begin{document}


\title{
Scalarized black holes in the presence of the coupling to Gauss-Bonnet gravity
}

\author{Masato Minamitsuji and Taishi Ikeda}
\affiliation{Centro de Astrof\'{\i}sica e Gravita\c c\~ao  - CENTRA,
Departamento de F\'{\i}sica, Instituto Superior T\'ecnico - IST,
Universidade de Lisboa - UL, Av. Rovisco Pais 1, 1049-001 Lisboa, Portugal}

\begin{abstract}
In this paper, we study static and spherically symmetric black hole (BH) solutions in the scalar-tensor theories with the coupling of the scalar field to the Gauss-Bonnet (GB) term $\xi (\phi) R_{\rm GB}$, where $R_{\rm GB}:=R^2-4R^{\alpha\beta}R_{\alpha\beta}+R^{\alpha\beta\mu\nu}R_{\alpha\beta\mu\nu}$ is the GB invariant and $\xi(\phi)$ is a function of the scalar field $\phi$. Recently, it was shown that in these theories scalarized static and spherically symmetric BH solutions which are different from the Schwarzschild solution and possess the nontrivial profiles of the scalar field can be realized for certain choices of the coupling functions and parameters. These scalarized BH solutions are classified in terms of the number of nodes of the scalar field. It was then pointed out that in the case of the pure quadratic order coupling to the GB term, $\xi(\phi)=\eta \phi^2/8$, scalarized BH solutions with any number of nodes are unstable against the radial perturbation. In order to see how a higher order power of $\phi$ in the coupling function $\xi(\phi)$ affects the properties of the scalarized BHs and their stability, we investigate scalarized BH solutions in the presence of the quartic order term in the GB coupling function, $\xi(\phi)=\eta \phi^2 (1+\alpha \phi^2)/8$. We clarify that the existence of the higher order term in the coupling function can realize scalarized BHs with zero nodes of the scalar field which are stable against the radial perturbation.
\end{abstract}

\maketitle  


\section{Introduction}
\label{sec1}

\subsection{Black holes in the scalar-tensor theories}
\label{sec11}

Scalar fields ubiquitously appear in the various contexts of theoretical physics.
The Lovelock theorem \cite{Lovelock:1972vz} states
that 
in four dimensions 
any rank-2, symmetric, and divergence-free tensor $A^{\mu\nu}$,
which depends on the metric tensor $g^{\mu\nu}$ and its first order derivatives
and linearly on its second order derivatives,
is given by a combination of the Einstein tensor $G^{\mu\nu}$
and $g^{\mu\nu}$ itself, 
$A^{\mu\nu}= p G^{\mu\nu}+ q g^{\mu\nu}$ with $p$ and $q$ being constants.
The theorem shows
that the only diffeomorphism invariant gravitational theory,
which is composed of only the metric tensor and its derivatives,
is general relativity (GR) with or without a cosmological constant. 
In other words,
any violation of the assumptions in the Lovelock theorem,
such as 
(i) fields other than the metric,
(ii) dimensionality of spacetime  larger than four,
(iii) existence of more than second order derivatives of the metric,
and/or 
(iv) nonlocality,
lead to gravitational theories 
which possess scalar and/or vector field degrees of freedom
after suitable mathematical transformations 
such as dimensional reduction and conformal/disformal transformations,
namely 
scalar-tensor (or vector-tensor) theories \cite{Berti:2015itd}.

The necessity of scalar fields has arisen
not only theoretically but also observationally.
The latest issues in cosmology
about the existence of primordial inflation and the late-time acceleration of the Universe
have motivated us 
to introduce some new sectors
beyond the conventional framework of GR and/or the Standard Model of particle physics.
Scalar fields
have provided the simplest but most convenient frameworks
which could realize accelerated expansion of the early or late-time Universe.
On the other hand,
the suppression of extra scalar forces
for the optimistic choice of parameters
requires some screening mechanisms \cite{Koyama:2015vza},
which exist beyond the conventional scheme of the scalar-tensor theories.
Studies in this direction have rediscovered 
the Horndeski theories
\cite{Horndeski:1974wa,Deffayet:2009mn,Deffayet:2011gz,Kobayashi:2011nu},
known as the most general scalar-tensor theories
with the second order equations of motion.
Moreover,
the scalar-tensor theories to explain 
the accelerated expansion of the Universe 
have been extended to the beyond Horndeski theories
\cite{Gleyzes:2014dya,Langlois:2015cwa,BenAchour:2016fzp}.

Besides studies of cosmology and phenomenology,
physics of black holes (BHs) 
will also be able to probe 
the existence of scalar fields from the completely different aspects.
In strong gravity regimes in/around BHs and relativistic stars,
nonlinearities of the gravitational field equations 
become most efficient
and 
the deviations from GR
will appear most significantly~\cite{Berti:2015itd,Barack:2018yly}.
One of the criteria 
for classifying the scalar-tensor theories in terms of BH physics
is the existence of the no-hair theorem for asymptotically flat BH solutions.
It has been shown that 
the BH no-hair theorem
holds for 
the scalar-tensor theories 
with canonical or noncanonical kinetic term, 
non-negative potential, 
and nonminimal coupling of the scalar field to the scalar curvature
(See e.g., Refs. \cite{Israel:1967wq,Carter:1971zc,Ruffini:1971bza,Hawking:1971vc,Chase,Hawking:1972qk,Bekenstein:1995un,Graham:2014mda,Graham:2014ina,Herdeiro:2015waa}.),
and the shift symmetric subclass of the (beyond) Horndeski theories
with the regular coupling functions of
the scalar field $\phi$ and the canonical kinetic term
$X:=-(1/2)g^{\mu\nu}\partial_\mu\phi \partial_\nu\phi$~\cite{Hui:2012qt,Babichev:2017guv}.
For the class of scalar-tensor theories 
which share the same solutions with GR and satisfy all the conditions for the no-hair theorem,
the Schwarzschild and Kerr solutions 
are the unique solutions of the gravitational field equations
in the static and spherically symmetric, and the stationary and axisymmetric cases, 
respectively.
Thus,
in order to test the existence of the scalar field degree of freedom,
perturbative properties of BHs have to be investigated,
since
the behavior of the perturbations may be different,
even if GR and scalar-tensor theories share the same BH solutions \cite{Barausse:2008xv,Tattersall:2018nve}.

On the other hand,
in the scalar-tensor theories,
where some of the conditions for the no-hair theorem are violated,
BH solutions different from the Schwarzschild and Kerr solutions
may exist.
Such a BH would be equipped with a nontrivial profile of the scalar field 
outside the event horizon.
One of the scalar-tensor theories which admit BH solutions absent in GR
is the scalar-tensor theory
with the coupling of the scalar field to the Gauss-Bonnet (GB) term
$\xi(\phi)R_{\rm GB}$ in the Lagrangian density,
where 
$\xi(\phi)$ is an arbitrary regular function of the scalar field,
and 
\begin{align}
R_{\rm GB}:=
R^2-4R^{\alpha\beta}R_{\alpha\beta}
+R^{\alpha\beta\mu\nu}R_{\alpha\beta\mu\nu},
\end{align}
is the GB term (See Eq.~\eqref{esgb} below.).
BH solutions in the scalar-tensor theories with the coupling to the GB term
have been studied since 1990s.
The first studies have focused on 
exponential type couplings $\xi(\phi)\propto e^{-c\phi}$
with $c$ being constant,
which is motivated by string theory.
\footnote{
Although a constant term may be added to the coupling function $\xi(\phi)$,
since it does not contribute to the equations of motion in four dimensions,
we will not consider it in the rest of the paper.} 
The static and spherically symmetric BH solutions 
have been investigated in
Refs. 
\cite{Kanti:1995vq,Alexeev:1996vs,Torii:1996yi,Kanti:1997br,Chen:2006ge,Guo:2008hf,Guo:2008eq,Ohta:2009tb,Ohta:2009pe,Lee:2018zym}
and then
generalized to
slowly rotating \cite{Pani:2009wy,Ayzenberg:2014aka,Maselli:2015tta}
and 
full rotating \cite{Kleihaus:2011tg,Kleihaus:2015aje} solutions accordingly.

The linear coupling of the scalar field to the GB term
$\xi(\phi)=c' \phi$ with $c'$ being constant
has also been considered,
since it corresponds 
to a class of the shift-symmetric Horndeski theories 
with the logarithmic dependence of the coupling functions on $X$,
which does not meet one of the conditions for the BH no-hair theorem~\cite{Hui:2012qt}.
Static and spherically symmetric
BH solutions in the presence of the linear coupling of the scalar field to the GB term
have been explicitly constructed in Refs. \cite{Sotiriou:2013qea,Sotiriou:2014pfa}.
They considered the expansion of the equations of motion
in terms of the small dimensionless coupling constant $\hat c:= c'/r_h^2\ll 1$,
where $r_h$ is the radial position of the event horizon.
At the order of ${\hat c}$,
the scalar field is sourced by the GB term 
of the Schwarzschild spacetime,
and 
at the order of ${\hat c}^2$,
the deviation from the Schwarzschild geometry
appears due to the backreaction of the scalar field.
The regularity of the scalar field at the horizon $r=r_h$ 
relates the mass with the scalar charge,
and, hence, the scalar charge is not independent of the mass.

It is important to emphasize that
if one considers the coupling function $\xi(\phi)$
which is the monotonic function of $\phi$,
such as $\xi=e^{c\phi}$ and $\xi= c'\phi$,
there can be only the nontrivial BH solutions,
while the Schwarzschild and Kerr metrics are {\it not} solutions.
On the other hand,  
if the coupling function $\xi(\phi)$ has at least one extremum at $\phi=\phi_0$,
$\xi^{(1)}(\phi_0)=0$
where $\xi^{(n)} (\phi):= \partial^n \xi (\phi)/\partial \phi^n$,
there can exist the Schwarzschild and Kerr solutions 
with $\phi=\phi_0$ as a solution,
as discussed in Sec. \ref{sec12}.

\subsection{Spontaneous scalarization}
\label{sec12}

In this paper,
we consider the scalar-tensor theory 
which is composed of 
the Einstein-Hilbert term,
the canonical kinetic term,
the coupling of the scalar field to the GB term,
and 
the matter field directly coupled to the scalar field:
\begin{align}
\label{esgb}
S=
\frac{1}{2\kappa^2}
\int d^4 x\sqrt{-g}
\left(
R 
-\frac{1}{2}g^{\mu\nu}\partial_\mu\phi \partial_\nu\phi
+ \xi (\phi)R_{\rm GB}
\right)
+\int d^4x \sqrt{-{\tilde g}}
  {\cal L}_m
    \left[
     {\tilde g}_{\mu\nu},\Psi
    \right],
\end{align}
where
$\kappa^2=8\pi G$ with $G$ being the gravitational constant,
the Greek indices $\mu,\nu,\cdots$ run over the space and time of the four-dimensional spacetime,
$g_{\mu\nu}$ is the metric tensor,
$g={\rm det}(g_{\mu\nu})$ is its determinant,
$R$ is the scalar curvature associated with $g_{\mu\nu}$,
$\phi$ is the scalar field,
and
${\cal L}_m$ is the Lagrangian for the matter field $\Psi$.
The Jordan frame metric ${\tilde g}_{\mu\nu}$
is  related to the Einstein frame one
by ${\tilde g}_{\mu\nu} =e^{2A(\phi)} g_{\mu\nu}$,
where $A(\phi)$ is the conformal factor depending on the scalar field only.
Throughout the paper,  we set $c=\hbar=1$.

Varying the action \eqref{esgb}
with respect to the metric $g_{\mu\nu}$,
the gravitational equation of motion is given by  
\cite{Guo:2008hf,Guo:2008eq,Ohta:2009tb,Ohta:2009pe}
\begin{align}
\label{cov_grav_eq}
  G_{\mu\nu}
=
\frac{1}{2}
  \left(
    \nabla_\mu \phi \nabla_\nu \phi 
-\frac{1}{2} g_{\mu\nu}\nabla^\lambda \phi \nabla_\lambda \phi 
  \right)
-4\left(\nabla^\rho\nabla^\sigma\xi (\phi)\right) P_{\mu\rho\nu\sigma}
+\kappa^2 {T}_{\mu\nu},
\end{align}
where 
\begin{align}
P_{\mu\nu\rho\sigma}
&:=
R_{\mu\nu\rho\sigma}
+g_{\mu\sigma}R_{\rho\nu}
-g_{\mu\rho}R_{\sigma\nu}
+g_{\nu\rho}R_{\sigma\mu}
-g_{\nu\sigma}R_{\rho\mu}
+\frac{R}{2}
\left(
  g_{\mu\rho}g_{\nu\sigma}
-  g_{\mu\sigma}g_{\nu\rho}
\right).
\end{align}
The matter energy-momentum tensors in the Jordan and Einstein frames,
respectively,
are defined by 
\begin{align}
{\tilde T}_{\mu\nu}
:=- \frac{2}{\sqrt{-{\tilde g}}}
   \frac{\delta \left(\sqrt{-{\tilde g}} {\cal L}_m[{\tilde g}_{\mu\nu} ,\Psi]\right)}
         {\delta {\tilde g}^{\mu\nu}},
\qquad 
 T_{\mu\nu}
:= - \frac{2}{\sqrt{-{ g}}}
 \frac{\delta \left(\sqrt{-{\tilde g}} {\cal L}_m[{\tilde g}_{\mu\nu},\Psi]\right)}
         {\delta { g}^{\mu\nu}},
\end{align}
which are related by 
\begin{align}
{\tilde T}_{\mu\nu}
= \sqrt{\frac{g}{\tilde g}}
  \frac{\delta g^{\alpha\beta}}{\delta {\tilde g}^{\mu\nu}}
  T_{\alpha\beta}
= e^{-2A(\phi)}
  T_{\mu\nu}.
\end{align}
On the other hand,
varying the action \eqref{esgb}
with respect to the scalar field $\phi$,
the equation of motion for the scalar field is given by 
\begin{align}
\label{cov_sca_eq}
\Box\phi+\xi^{(1)}(\phi)R_{\rm GB}
+2\kappa^2 A^{(1)}(\phi) T^\mu{}_\mu=0,
\end{align}
where $\Box\phi=g^{\mu\nu} \nabla_\mu\nabla_\nu\phi$
is the d'Alembertian operator
and  
$A^{(n)}(\phi) := \partial^n A(\phi)/\partial \phi^n$.

In the case that 
$\xi(\phi)$ and $A(\phi)$ satisfy $\xi^{(1)}(\phi_0)=A^{(1)}(\phi_0)=0$
at $\phi=\phi_0={\rm constant}$,
where the scalar field equation of motion \eqref{cov_sca_eq}
is trivially satisfied,
since
$\nabla_\rho\nabla_\sigma \xi
=\partial_\rho\partial_\sigma \xi-\Gamma_{\rho\sigma}^\alpha \partial_\alpha \xi
=\xi^{(1)} \partial_\rho \partial_\sigma \phi
+\xi^{(2)} \partial_\rho \phi \partial_\sigma \phi
-\Gamma_{\rho\sigma}^\alpha \xi^{(1)} \partial_\alpha \phi
=0$ at $\phi=\phi_0$,
the gravitational equation of motion \eqref{cov_grav_eq}
reduces to the GR one:
\begin{align}
\label{gr}
G_{\mu\nu}=
\kappa^2 T_{\mu\nu},
\end{align}
where we assumed that $A(\phi_0)=0$.
Thus, 
the metric $g_{\mu\nu}$ satisfies the Einstein equation \eqref{gr}
and 
the constant scalar field $\phi=\phi_0$,
such that $\xi^{(1)}(\phi_0)=A^{(1)}(\phi_0)=0$
is a solution of the theory \eqref{esgb}.
For instance,
if $\xi (\phi)\propto \phi^2$ and $A(\phi)\propto \phi^2$,
a solution in GR with $\phi=0$ is a solution. 
Nevertheless,
this does not guarantee the uniqueness 
of the GR solution with a constant scalar field 
and
there may be another solution 
with a nontrivial profile of the scalar field.
If the theory \eqref{esgb} admits two or more solutions
and
if one of them is the GR solution with a constant scalar field, 
an interesting question is 
which solution is dynamically favored?
If the GR solution is unstable against the linear perturbation 
in a local region of the spacetime,
the so-called spontaneous scalarization may take place.

Spontaneous scalarization has been argued
for relativistic stars
in the scalar-tensor theory 
with the coupling of the scalar field to the matter field $A(\phi)\neq 0$
[and without the coupling to the GB term $\xi(\phi)=0$],
which is triggered by a tachyonic instability of the scalar field
due to the coupling to the matter field.
In the simplest realization
discussed by Damour and Esposito-Far\`ese
\cite{Damour:1993hw,Damour:1996ke},
the scalarization occurs for the coupling $A(\phi)=\beta \phi^2/8$,
where $\beta$ is a constant.
The presence of the scalar field can significantly
modify the properties of relativistic stars, 
while satisfying the weak field constraints.
Irrespective of the choice of the equation of state,
the scalarization occurs for $\beta\lesssim -4.35$
\cite{Novak:1997hw,Harada:1998ge},
while
the binary-pulsar observations 
have put the bounds $\beta\gtrsim -4.5$ \cite{Freire:2012mg}.

\subsection{Gravitational spontaneous scalarization: Test field analysis}
\label{sec13}

\subsubsection{Tachyonic instability of the Schwarzschild solution}
\label{sec131}

Following the structure of Eq.~\eqref{cov_sca_eq},
we expect that
spontaneous scalarization
may take place
even in a vacuum spacetime with ${\cal L}_{m}=0$ in Eq.~\eqref{esgb}.
In the case of the spherically symmetric spacetime,
a tachyonic instability of the vacuum Schwarzschild solution
\begin{align}
\label{sch}
ds^2=
-\left(1-\frac{2M}{r}\right)dt^2
+\frac{dr^2}{1-2M/r}
+r^2 \left(d\theta^2+\sin^2\theta d\varphi^2\right),
\end{align}
could be triggered due to the coupling of the scalar field to the GB term $\xi (\phi)\neq 0$,
where 
$r$ and $M$ are the radial coordinate and mass of the Schwarzschild spacetime,
respectively.

The simplest case that the scalar-tensor theory \eqref{esgb} admits the Schwarzschild solution \eqref{sch}
is the case of the quadratic order coupling of the scalar field to the GB term~\cite{Silva:2017uqg}:
\begin{align}
\label{quadratic}
\xi(\phi)=\frac{\eta}{8}\ \phi^2,
\end{align}
where $\eta(>0)$ is a constant with mass dimension $(-2)$,
for which
the metric \eqref{sch} with $\phi=0$ is a solution to the theory \eqref{esgb}.
In general, the Schwarzschild solution with $\phi=0$ can be realized 
for a more general coupling function with the higher order powers of the scalar field:
\begin{align}
\label{higherorder}
\xi(\phi)=\frac{\eta}{8}\ \phi^2 +\sum_{i\geq 3}c_i \phi^i,
\end{align}
where $c_i$ ($i=3,4,5,\cdots$) are constants with mass dimension $(-2)$.

For the coupling \eqref{higherorder},
we then consider the small perturbation of the scalar field $\delta\phi (x^\mu)$
about the Schwarzschild solution \eqref{sch} with $\phi=0$.
By neglecting the metric perturbations, 
$\delta\phi (x^\mu)$ obeys the equation
\begin{align}
\label{perturb0}
  \Box_{\rm Sch}\delta \phi 
+\xi^{(1)}(\delta \phi)R_{\rm GB}^{({\rm Sch})}
=
\left(
  \Box_{\rm Sch}
+\frac{\eta}{4}  R_{\rm GB}^{({\rm Sch})}
\right)
\delta \phi
+{\cal O}(\delta\phi^2)
=0,
\end{align}
where $R_{\rm GB}^{({\rm Sch})}=48M^2/r^6$
is the GB term for the Schwarzschild solution \eqref{sch}
and $\Box_{\rm Sch} $ is the d'Alembertian operator
on the Schwarzschild background \eqref{sch}.
Decomposing $\delta\phi (x^\mu)$ into the partial modes:
\begin{align}
\label{fourier}
\delta\phi(x^\mu)=
\sum_{\ell,m}
 e^{-i\omega_{\ell m} t} \frac{\sigma_{\ell m} (r)}{r}Y_{\ell m} (\Omega),
\end{align}
where $Y_{\ell m}(\Omega)$ is the spherical harmonics,
from Eq.~\eqref{perturb0}
the radial mode equation is given by
\begin{align}
\label{radial}
\left[
-\frac{d^2}{dr_\ast^2}
+V_{\rm eff}(r)
\right]
\sigma_{\ell m}(r)
=\omega_{\ell m}^2
\sigma_{\ell m}(r),
\end{align}
where
$dr_\ast:=dr/(1-2M/r)$ is the tortoise radial coordinate for 
the Schwarzschild spacetime
and  the effective potential is given by 
\begin{align}
\label{ep}
V_{\rm eff}(r)
=
(r-2M)
\left[
\frac{2M}{r^7} (r^3-6M\eta)
+
\frac{\ell (\ell+1)}{r^3}
\right].
\end{align}
For the radial perturbation ($\ell=0$),
in the vicinity of the horizon $r=2M$,
the effective potential \eqref{ep} 
can be approximated by $V_{\rm eff} (r)\simeq (4M^2-3\eta)(r-2M) /(32M^5)$,
and hence 
$V_{\rm eff}(r)$
contains a negative region for $\eta/M^2>4/3\approx 1.333$,
implying
the appearance of pure imaginary modes,
i.e., the modes with imaginary $\omega$.
On the other hand,
for $\eta/M^2<4/3$
there is no negative region in the effective potential,
and, hence,
the Schwarzschild solution \eqref{sch} with $\phi=0$
is expected to be linearly stable for the radial perturbation.
From Eq. \eqref{ep},
we also find that, if the effective potential $V_{\rm eff}(r)$ for the radial perturbation ($\ell=0$)
is non-negative,
$V_{\rm eff}(r)$
for the nonradial perturbation ($\ell\geq 1$)
is also non-negative.

\subsubsection{Test scalar field solutions: Pure quadratic coupling}
\label{sec132}

In order to check the existence of the scalarized BH solutions
in the case without the matter field ${\cal L}_m=0$ in a semiquantitative way,
we present static solutions of the test scalar field
by neglecting the backreaction on the spacetime geometry.
We will solve the scalar field equation of motion \eqref{cov_sca_eq}
under the static and spherically symmetric ansatz 
$\phi=\psi(r)$
on a fixed Schwarzschild background \eqref{sch}. 
For a GB coupling function $\xi(\phi)$,
the scalar field equation of motion Eq.~\eqref{cov_sca_eq} reduces to 
\begin{align}
\label{testeq}
(r-2M)
\psi'' (r)
+\frac{2(r-M)}{r}\psi' (r)
+
\frac{48M^2}{r^5}
\xi^{(1)}(\psi) 
=0,
\end{align}
where a ``prime'' denotes the derivative 
with respect to the radial coordinate $r$. 
The general regular solution of the static scalar field $\phi=\psi(r)$  in the vicinity of the event horizon $r=2M$ is given by 
\begin{align}
\label{psi_horizon}
\psi(r)
=\psi_0
-\frac{3\xi^{(1)} (\psi_0)}{2M^3} (r-2M)
+{\cal O} \left((r-2M)^2\right),
\end{align}
where
$\psi_0$ is a constant,
which represents the value of the scalar field at the event horizon and can take an arbitrary nonzero value,
and we have assumed the regularity of the coupling function $\xi(\phi)$. 
We note that besides Eq.  \eqref{psi_horizon}
there is the second solution of Eq. \eqref{testeq},
which is singular at the horizon and, hence, not considered here.
Similarly, 
the solution for the scalar field at the asymptotic spatial infinity $r\to \infty$ is given by  
\begin{align}
\psi(r)
=\psi_\infty
+\frac{Q}{r}
+\frac{MQ}{r^2}
+\frac{4M^2Q}{3r^3}
+\frac{2M^2(MQ-2\xi^{(1)}(\psi_\infty))}{r^4}
+{\cal O} \left(\frac{1}{r^5}\right),
\end{align}
where $\psi_\infty$ and $Q$ are constants,
which represent the values of the scalar field in the limit of $r\to \infty$ 
and the scalar charge, respectively.

We numerically solve Eq.~\eqref{testeq} 
from a point sufficiently close to the horizon toward the region of large $r$,
under the boundary conditions obtained from Eq.  \eqref{psi_horizon}:
\begin{align}
\psi(r\to 2M)=\psi_0\neq 0,
\qquad
\psi'(r\to 2M)=-\frac{3\xi^{(1)} (\psi_0)}{2M^3}.
\end{align}
We also choose the boundary condition $\psi_\infty=0$ at the spatial infinity,
as the scalarization of BHs is triggered
by a tachyonic instability of the trivial solution $\psi=0$.
For $\psi_0=0$, there is only the trivial solution $\psi=0$.
On the other hand,
for $\psi_0\neq 0$ there will be a nontrivial scalar field solution $\psi(r)\neq 0$. 
The scalar charge $Q$ is read from the behavior of the scalar field at the spatial infinity
\begin{align}
\label{scalar_charge_test}
Q=-r^2\psi'(r)\Big|_{r\to \infty}.
\end{align}
For our numerical analysis, we set $M=1$
and integrate \eqref{testeq}
from $r=2M+10^{-5}$ to $r=1.0\times 10^5$.
The scalar charge $Q$ is then numerically evaluated 
via Eq. \eqref{scalar_charge_test} at $r=1.0\times 10^5$.

The small perturbation about the static scalar field solution,
$\phi=\psi(r)+\delta\phi (x^\mu)$,
obeys the equation
\begin{align}
\label{perturb1}
\left(
  \Box_{\rm Sch} 
-m_{\rm eff}^2
\right)
\delta\phi
=0,
\end{align}
where 
$m_{\rm eff}^2:=-\xi^{(2)}(\psi)  R_{\rm GB}^{({\rm Sch})}$
is the effective mass squared of the scalar field perturbation.
Decomposing $\delta\phi$ as Eq. \eqref{fourier}
and deriving the radial mode equation as Eq. \eqref{radial}
with the tortoise coordinate of the Schwarzschild spacetime $r_\ast$,
the effective potential 
for each mode of the linear perturbation $\delta\phi (x^\mu)$ is then given by 
\begin{align}
\label{ep2}
V_{\rm eff}(r)
=
(r-2M)
\left[
\frac{2M}{r^7} 
\left(r^3-24 M \xi^{(2)} (\psi)\right)
+
\frac{\ell (\ell+1)}{r^3}
\right].
\end{align}

\begin{figure}[h]
\unitlength=1.1mm
\begin{center}
  \includegraphics[height=7.5cm,angle=0]{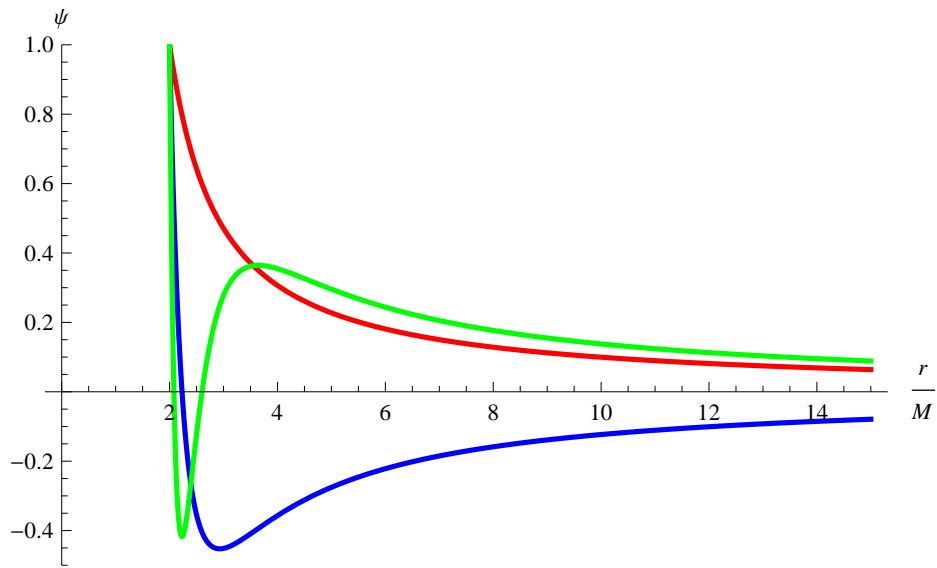}
\caption{
The solutions $\psi(r)$ to Eq. \eqref{testeq} for the quadratic order coupling to the GB term \eqref{quadratic}
with the boundary conditions $\psi_0=1.0$ and $\psi_\infty=0$
are shown
as the function of $r/M$.
The red, blue, and green curves correspond 
to the scalar field solutions  
with zero, one, and two nodes of the scalar field, respectively.
}
  \label{figtest}
\end{center}
\end{figure}

For the quadratic order coupling~\eqref{quadratic} considered in Sec. \ref{sec131},
as shown in Fig.~\ref{figtest},
the scalar field solutions of Eq.~\eqref{testeq} with zero, one and two nodes 
can be numerically obtained 
for $\eta/M^2=2.90$, $19.5$, $50.9$ \cite{Silva:2017uqg},
respectively.
\footnote{ 
There will be solutions of the scalar field with more nodes,
but in this paper
we focus on the scalar field solutions with zero, one, and two nodes.}
The scalar charge $Q$ is given by 
\begin{align}
\label{charge}
\frac{Q}{\sqrt{\eta}}
=0.525 \psi_0, 
\,\,
- 0.250 \psi_0,
\,\,
 0.174\psi_0,
\end{align}
for the solutions with zero, one, and two nodes.
For the quadratic order coupling to the GB term~\eqref{quadratic},
from Eq.~\eqref{perturb1}
we find that 
the equation for the small perturbation $\phi=\psi (r)+\delta \phi(x^\mu)$
follows the same equation as the leading order part of Eq.~\eqref{perturb0}.
Since all the values $\eta/M^2=2.90$, $19.5$, $50.9$, for the nontrivial solutions of the scalar field
are larger than $4/3$,
the effective potential \eqref{ep} for the radial perturbation ($\ell=0$)
contains a negative region
and would possess pure imaginary modes.  
Thus, 
we expect that the nontrivial scalar field solutions for Eq.~\eqref{quadratic}
are unstable against the radial perturbation.

By taking the backreaction on the spacetime geometry,
the explicit solutions of scalarized BH solutions
were firstly presented in Refs.  \cite{Silva:2017uqg,Doneva:2017bvd,Antoniou:2017acq,Antoniou:2017hxj,Kleihaus:2015aje},
and then followed by Refs. ~\cite{Doneva:2017duq,Blazquez-Salcedo:2018jnn,Doneva:2018rou}.
In Refs. \cite{Silva:2017uqg} and \cite{Doneva:2017bvd},
the scalarized BH solutions
for the coupling functions \eqref{quadratic}
and 
$\xi(\phi)= \eta(1- e^{-3\phi^2/2})/12$ 
in our conventions
were investigated, respectively.
Ref. \cite{Blazquez-Salcedo:2018jnn}
analyzed the radial perturbation of the scalarized BH solutions
for both the coupling functions.
They confirmed the existence of pure imaginary modes for all the scalarized BHs
for the quadratic order coupling \eqref{quadratic}.
This instability can be qualitatively understood
from our test field analysis shown above.
On the other hand,
for the coupling function 
$\xi(\phi)= \eta(1- e^{-3\phi^2/2})/12$,
Ref. \cite{Blazquez-Salcedo:2018jnn} also showed that
the effective potential for scalarized BH solutions with zero nodes
whose mass is greater than the critical mass
contains no pure imaginary mode,
and, hence, these solutions are radially stable. 
These results indicate
the importance of the existence of the higher order powers of $\phi$ 
in the coupling constant $\xi(\phi)$
to realize scalarized BH solutions which are stable 
against the radial perturbation. 
In order to clarify this point, 
we will study the coupling function $\xi(\phi)$
which contains the $\phi^4$ term as well as the $\phi^2$ term.

\subsubsection{Test scalar field solutions: General coupling}
\label{sec133}
\begin{figure}[h]
\unitlength=1.1mm
\begin{center}
  \includegraphics[height=7.5cm,angle=0]{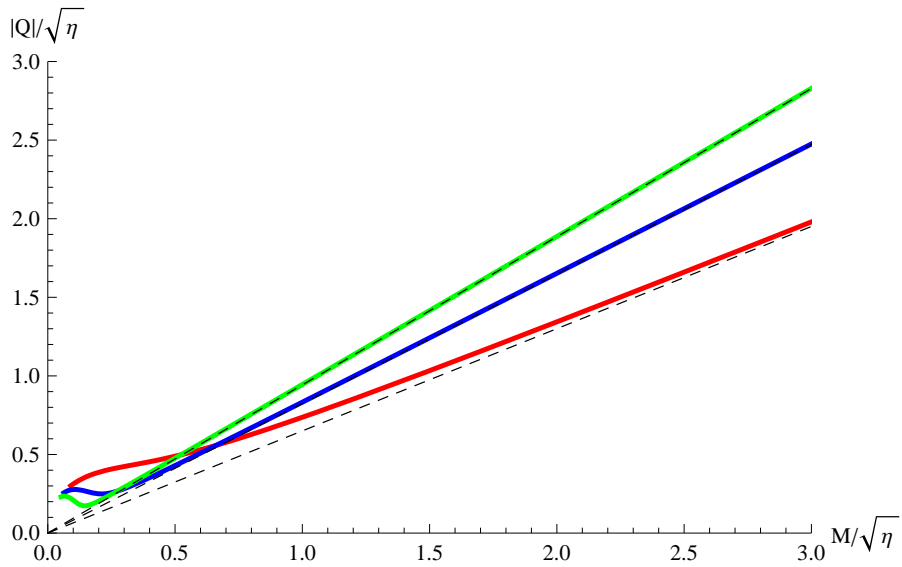}
\caption{
The absolute value of the scalar charge $|Q|$ divided by $\sqrt{\eta}$
is shown as the function of the mass $M$ divided by $\sqrt{\eta}$
for $\psi_0=1.0$.
We note that $Q>0$ for the scalar field solutions with an even number of nodes,
and $Q<0$ for those with an odd number of nodes.
The red, blue, and green curves
correspond to 
the scalar field solutions 
with zero, one, and two nodes,
respectively.
The dashed lines from the bottom
correspond to 
the solutions with zero, one, and two nodes
for the pure quartic order coupling $\xi (\phi)=\lambda \phi^4/8$,
respectively.
For $\alpha<0$,
we have chosen the minimal value of $\alpha$ to $-0.5 + 10^{-7}$.
}
  \label{figtestgen}
\end{center}
\end{figure}

In this paper,
we will consider the GB coupling function: 
\begin{align}
\label{general}
\xi(\phi)=
\frac{\eta}{8}\left( \phi^2+\alpha \phi^4 \right),
\end{align}
which preserves the symmetry $\xi(-\phi)=\xi(\phi)$,
where $\alpha$ is the dimensionless coupling constant.
For the coupling function \eqref{general}, Eq.~\eqref{testeq} can be rewritten by   
\begin{align}
\label{testq2}
 \psi''(r)
+\frac{2(r-M)}{r (r-2M)} \psi' (r)
+\frac{12\eta M^2}{r^5 (r-2M)} \psi (r)
  \left(1+2 \alpha \psi(r)^2\right)
=0.
\end{align}
There are several solutions of a constant scalar field:
For $\alpha>0$, $\psi=0$ is only the constant scalar field solution,
while for $\alpha<0$, in addition to $\psi=0$,
there are two additional constant scalar field solutions 
\begin{align}
\label{gr_sol}
\psi(r)=\pm \sqrt{-\frac{1}{2\alpha}}.
\end{align}
We note that 
even if the backreaction is taken into consideration
the Schwarzschild metric with these constant values of the scalar field
are the solutions of the theory \eqref{esgb}.
Since $R_{\rm GB}^{({\rm Sch})}=48M^2/r^6>0$
and $\xi^{(2)} (\pm \sqrt{-1/(2\alpha)})=-\eta/2<0$,
the effective mass squared for the perturbation $m_{\rm eff}^2= -\xi^{(2)}(\psi)R^{(Sch)}_{\rm GB}>0$
[See Eq. \eqref{perturb1}.];
the solution \eqref{gr_sol} 
is stable
and 
no spontaneous scalarization takes place.
On the other hand,  
for the solution with $\psi=0$,
$\xi^{(2)}(0)=\eta/4>0$,
and, hence, $m_{\rm eff}^2=-\xi^{(2)} (0)R_{\rm GB}<0$,
indicating the occurrence of a tachyonic instability.
In the limit of Eq. \eqref{testq2} to the event horizon $r\to 2M$,
assuming that the second order derivative $\psi'' (r)$ is regular,
the boundary condition at the horizon can be obtained as
\begin{align}
\psi' (r=2M)
=-\frac{3\eta\psi_0 }{8M^3}
    \left(1+2\alpha \psi_0^2\right).
\end{align}
Thus, for $\psi_0>0$,
the decreasing $\psi$ in the vicinity of the horizon 
imposes $\psi'(r=2M)<0$, and hence
\begin{align}
\label{general2_test}
\alpha  \psi_0^2>-\frac{1}{2}.
\end{align}
For $\psi_0<0$,
we impose $\psi'(r=2M)>0$, leading to the same condition as Eq.~\eqref{general2_test}.
For $\psi_0=\pm \sqrt{-1/(2\alpha)}$ with $\alpha<0$,
the solution is given by Eq.~\eqref{gr_sol},
which does not satisfy $\psi_\infty=0$.
Without loss of generality, we assume $\psi_0>0$. 
In Fig.~\ref{figtestgen},
the absolute value of the scalar charge $|Q|$ divided by $\sqrt{\eta}$
is shown as the function of the mass $M$ divided by $\sqrt{\eta}$
for $\psi_0=1.0$.
We note that $Q>0$ for the scalar field solutions with even number nodes,
and $Q<0$ for those with odd number nodes.
The red, blue, and green curves
correspond to 
the scalar field solutions 
with zero, one, and two nodes,
respectively.
The dashed lines from the bottom
correspond to 
the scalar field solutions with zero, one, and two nodes
for the pure quartic order coupling:
\begin{align}
\label{quartic}
\xi (\phi)=\frac{\lambda}{8} \phi^4,
\end{align}
which are given by 
$Q/M=0.650\psi_0, -0.824 \psi_0, 0.943\psi_0$, 
respectively,
where $\lambda$ is the coupling constant with mass dimension $(-2)$.
As we will see later,
the essential features remain the same 
even if the backreaction and the metric perturbation are included
in our analysis.

\begin{figure}[h]
\unitlength=1.1mm
\begin{center}
  \includegraphics[height=7.5cm,angle=0]{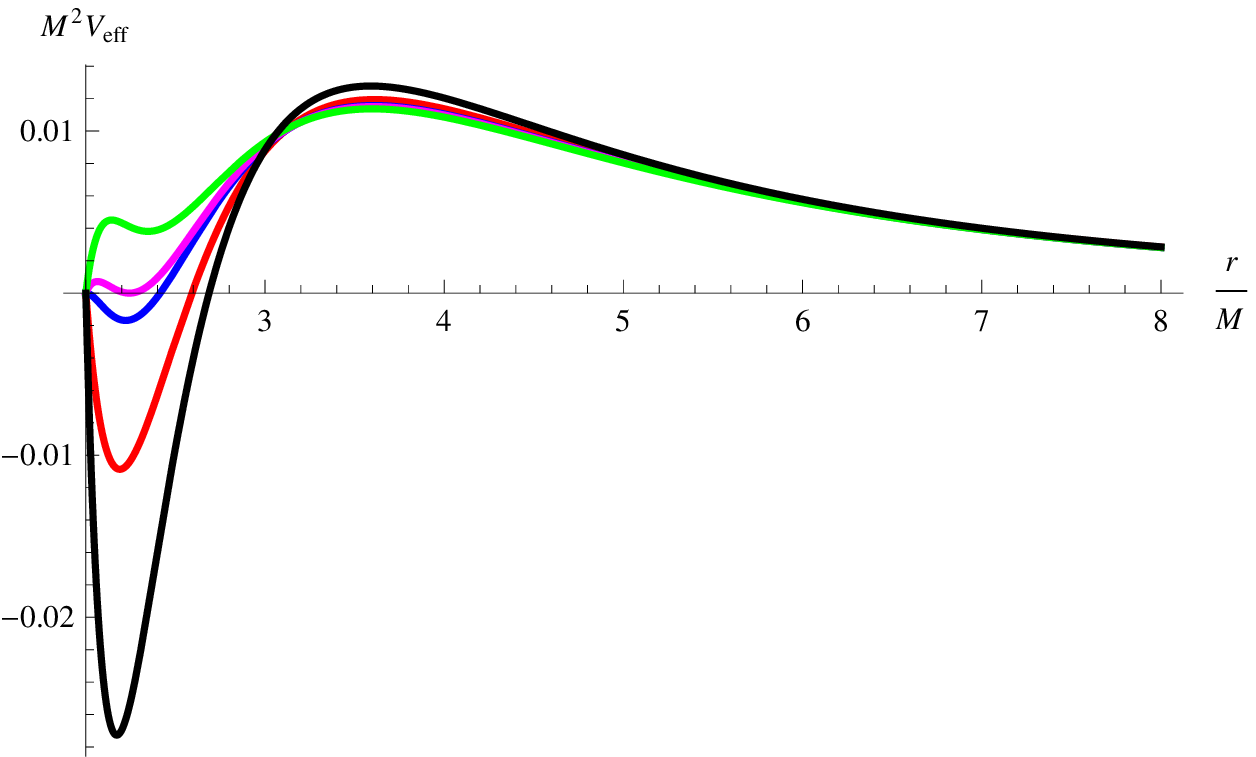}
  \includegraphics[height=5.0cm,angle=0]{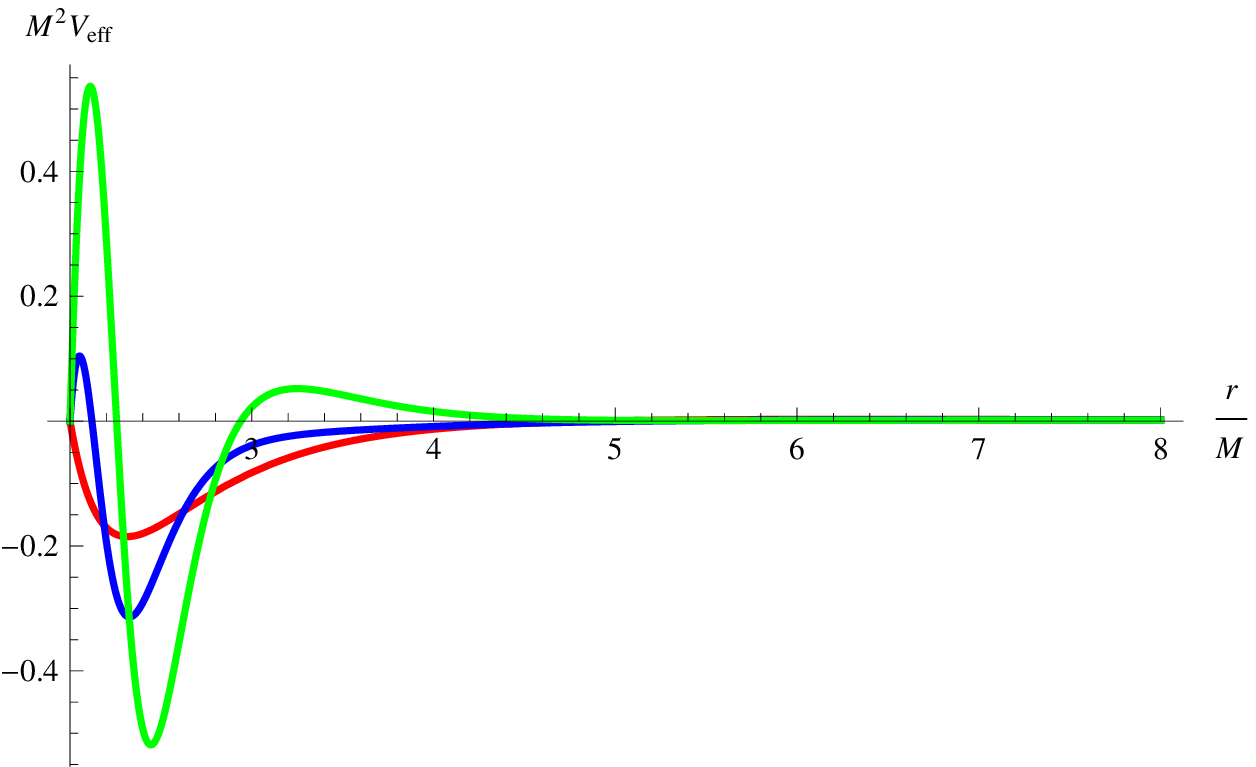}
  \includegraphics[height=5.0cm,angle=0]{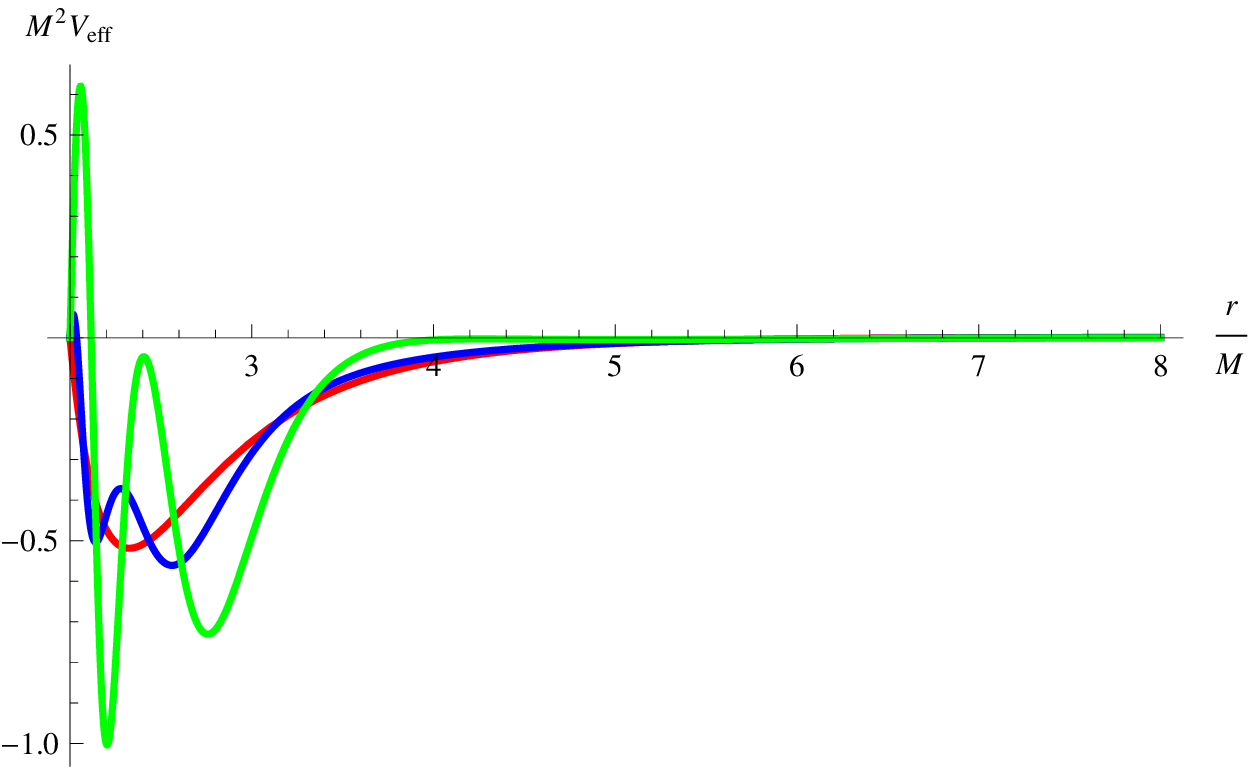}
\caption{
For the coupling \eqref{general},
the effective potential $V_{\rm eff} (r)$ multiplied by $M^2$
for the radial perturbation ($\ell=0$)
is shown as the function of $r/M$ for $\psi_0=1.0$
for the scalar field solutions with
zero, one, and two nodes,
respectively.
In the top panel,
the black, red, blue, magenta, and green curves
correspond to  
the scalar field solutions with zero nodes
for 
$(\eta/M^2,\alpha) =(2.14, 0.30)$, $(2.90,0)$, $(3.32,-0.10)$, $(3.402, -0.115)$,
$(3.59, -0.15)$,
respectively.
Similarly,
in the bottom-left panel,
the red, blue, and green curves
correspond to 
the scalar field solutions with one node
for $(\eta/M^2,\alpha)=(19.5,0)$, $(31.6,-0.40)$, $(53.5, -0.49)$,
and in the bottom-right panel,
the red, blue, and green curves
correspond to 
the scalar field solutions with two nodes
for $(\eta/M^2,\alpha)=(50.9,0)$, $(66.4,-0.35)$, $(105.9, -0.49)$,
respectively.
}
  \label{figeffpot}
\end{center}
\end{figure}

From Eq. \eqref{ep2}, 
for the coupling \eqref{general},
the effective potential 
for each mode of the perturbation $\phi=\psi(r)+\delta\phi (x^\mu)$
is given by 
\begin{align}
\label{ep3}
V_{\rm eff}(r)
=
(r-2M)
\left[
\frac{2M}{r^7} 
\left(r^3-6\eta M(1+6\alpha \psi^2)\right)
+
\frac{\ell (\ell+1)}{r^3}
\right].
\end{align}
In Fig.~\ref{figeffpot},
the effective potential $V_{\rm eff} (r)$ multiplied by $M^2$
for the radial perturbation ($\ell=0$)
is shown as the function of $r/M$ for $\psi_0=1.0$
for the scalar field solutions with zero, one, and two nodes,
respectively.
In the top panel,
the red, blue, magenta, and green curves
correspond to 
the scalar field solutions with zero nodes
for 
$(\eta/M^2,\alpha)= (2.14, 0.30)$, $(2.90,0)$, $(3.32,-0.10)$, $(3.402, -0.1155)$, $(3.59, -0.15)$,
respectively.
For $\alpha\psi_0^2=-0.1155$, 
we find that the local minimum of $V_{\rm eff}(r)$ 
for the scalar field solution
with zero nodes
 becomes zero,
and, hence, for 
\begin{align}
\label{alpha_bound}
-\frac{1}{2} <  \alpha \psi_0^2< -0.1155,
\end{align}
$V_{\rm eff}(r)$ has no negative region.
Thus, 
for such a solution, 
the spectrum does not contain any pure imaginary mode,
and the solution is expected to be stable against the radial perturbation ($\ell=0$).
From Eq. \eqref{ep3}, 
for $\alpha$ satisfying Eq. \eqref{alpha_bound}
the solutions of the static scalar field with zero nodes 
are also expected to be
stable against the nonradial perturbations ($\ell\geq 1$).

Similarly, 
in the bottom-left panel  of Fig. \ref{figeffpot},
the red, blue, and green curves
correspond to 
the scalar field solutions with one node
for 
$(\eta/M^2,\alpha)=(19.5,0)$, $(31.6,-0.40)$, $(53.5, -0.49)$,
and 
in the bottom-right panel of Fig. \ref{figeffpot},
the red, blue, and green curves
correspond to 
the scalar field solutions with two nodes
for 
$(\eta/M^2,\alpha)=(50.9,0)$, $(66.4,-0.35)$, $(105.9, -0.49)$,
respectively.
For the scalar field solutions with one and two nodes,
even for a large negative value of $\alpha$,
the effective potential always contains
a negative region, 
and hence 
should contain pure imaginary modes.
Thus, these solutions are expected to 
remain unstable against the radial perturbation.
We note that
for the constant scalar field solution Eq.~\eqref{gr_sol}
the effective potential \eqref{ep3} for the radial perturbation ($\ell=0$)
reduces to 
\begin{align}
\label{effpot_grsol}
V_{\rm eff}(r)= \frac{2M}{r^7}(r-2M) (r^3+12\eta M)>0,
\end{align}
which corresponds to the stable Schwarzschild solutions with a constant scalar field.
For the nonradial perturbations ($\ell \geq 1$), we also obtain $V_{\rm eff}(r)>0$,
which confirms the stability of the solution \eqref{gr_sol}.

Since the analysis in this subsection
does not include the backreaction on the spacetime geometry
and the metric perturbation,
the conclusion obtained from the above analysis
may change
if they are fully taken into consideration.
Nevertheless,
it is very suggestive that 
the test field analysis 
contains the stable scalar field solution with zero nodes
if there is a higher power of the scalar field
in the coupling function $\xi(\phi)$.
This motivates us 
to construct the scalarized BH solutions 
in the scalar-tensor theories \eqref{esgb}
by solving the gravitational equation of motion \eqref{cov_grav_eq}
and the scalar field equation of motion \eqref{cov_sca_eq},
and investigate their properties.
We will explore these issues in the rest of the paper.

\subsection{Construction of the paper}
\label{sec14}

We will organize the rest of the paper as follows:
In Sec. \ref{sec2}, 
we will present the strategy
for constructing the static and spherically symmetric vacuum BH solutions
in the scalar-tensor theories \eqref{esgb}. 
In Sec. \ref{sec3}, 
we will show 
the explicit solutions of the scalarized BH solutions
for the coupling functions \eqref{quadratic} and \eqref{general}.
In Sec. \ref{sec4}, 
we will investigate the stability
of the scalarized BH solutions constructed in Sec. \ref{sec3}
against the radial perturbation.
In Sec. \ref{sec5},
we will close the paper after giving a brief summary and conclusion.

\section{Strategy}
\label{sec2}

In the rest of the paper,
we will focus on the scalar-tensor theories 
\eqref{esgb}
without the matter field ${\cal L}_m=0$.

\subsection{Equations of motion}

We assume
that the metric is given by the form of a static and spherically symmetric spacetime:
\begin{align}
\label{static}
ds^2=
-A(r)dt^2+\frac{dr^2}{B(r)}
+r^2\left(d\theta^2+\sin^2\theta d\varphi^2\right),
\end{align}
where $A(r)$ and $B(r)$ are the functions of the radial coordinate $r$,
and the scalar field $\phi$ is also the function of $r$:  
\begin{align}
\phi=\psi(r).
\end{align}
The $(t,t)-$ and $(r,r)-$ components of the gravitational equation of motion \eqref{cov_grav_eq}
are then given by
\begin{align}
\label{eq1}
&2\left(r+2(1-3B)\xi^{(1)}\psi'\right)B' 
+\frac{1}{2}\left(-4+B(4+r^2\psi'^2)\right)
-8B(B-1)
\left(
\psi'^2 \xi^{(2)}+\psi''\xi^{(1)}  
\right)
=0,
\\
\label{eq2}
&2\left(r+2(1-3B)\xi^{(1)}\psi'\right)A' 
-\frac{A}{2B}
\left(
4+B (-4+r^2\psi'^2)
\right)
=0.
\end{align}
Similarly,
the scalar field equation of motion \eqref{cov_sca_eq}
is given by 
\begin{align}
\label{eq3}
\psi''
+\frac{1}{2}\left(\frac{4}{r}+\frac{A'}{A}+\frac{B'}{B}\right)\psi'
+\frac{4}{r^2A}(-1+B)\xi^{(1)} A''(r)
-\frac{2A' \xi^{(1)}}{r^2 A^2 B}
  \left(
     B^2 A'+AB'-A'B-3ABB'
   \right)
=0.
\end{align}
Solving Eq.~\eqref{eq2} with respect to $B^{-1}$,
we obtain
\begin{align}
\label{b}
\frac{1}{B}
=\frac{1}{8A}
\left(
4A' (r+2\xi^{(1)}\psi')+A (4-r^2\psi'^2)
\pm
\sqrt{
-384 A A' \xi^{(1)} \psi'
+
\left(
4A'(r+2\xi^{(1)}\psi')
+A(4-r^2\psi'^2) 
\right)^2
}
\right).
\end{align}
For a BH solution,
we require that for $r>r_h$,
where $r_h$ is the position of the event (outermost) horizon,
$A>0$, $A'>0$, $B>0$, $B'>0$,
and 
in the limit of $r\to r_h$,
$A\to 0$, $B\to 0$, $A/B \to {\rm constant}$,
and $\psi$ and $\psi'$ are regular.
Then, 
in the limit of $r\to r_h$,
Eq.~\eqref{eq2} can be approximated by 
\begin{align}
  \frac{1}{B}
\to 
\frac{A'}{A}
\left(
r+2\xi^{(1)} \psi'
\right)
\Big|_{r\to r_h},
\end{align}
which can be obtained only from the $(+)$-branch of Eq.~\eqref{b}.
Thus,
in the rest of the paper
we only focus on the $(+)$-branch.

Substituting Eq.~\eqref{b} into Eqs.~\eqref{eq1} and \eqref{eq3},
we obtain a set of the equations which are quasilinear for $A''(r)$ and $\psi'' (r)$,
and, by arranging them,
we obtain
the evolution equations for $A(r)$ and $\psi(r)$
with respect to $r$:
\begin{align}
\label{eq_A}
A''(r)&= F_A\left[A,A',\psi,\psi' ; r\right],
\\
\label{eq_psi}
\psi'' (r)&= F_\psi \left[A,A',\psi,\psi'; r\right],
\end{align}
where $F_A$ and $F_\psi$
are the nonlinear combinations of the given variables
which we omit to show explicitly,
since they are quite involved.
After integrating Eqs.~\eqref{eq_A} and \eqref{eq_psi}
for $A(r)$ and $\psi (r)$
and substituting them into Eq.~\eqref{b},
$B(r)$ can be obtained.

We also define
the right-hand side of the gravitational equation of motion \eqref{cov_grav_eq}
as the effective energy-momentum tensor of the scalar field:
\begin{align}
T^{\rm (eff)}_{\mu\nu}
:=
\frac{1}{2}
  \left(
    \nabla_\mu \phi \nabla_\nu \phi 
-\frac{1}{2} g_{\mu\nu}\nabla^\lambda \phi \nabla_\lambda \phi 
  \right)
-4\left(\nabla^\rho\nabla^\sigma\xi (\phi)\right) P_{\mu\rho\nu\sigma}.
\end{align}
In the static and spherically symmetric spacetime \eqref{static},
the effective energy density,
the radial pressure,
and 
the tangential pressure, defined by 
$\rho^{\rm (eff)}(r):
=-T^{\rm (eff)}{}^t{}_t$,
$p_r^{\rm (eff)}(r) 
:=T^{\rm (eff)}{}^r{}_r$,
and 
$p_t^{\rm (eff)}(r)
:=T^{\rm (eff)}{}^\theta{}_\theta
[=T^{\rm (eff)}{}^\varphi{}_\varphi]$,
respectively,
are explicitly given by  
\begin{align}
\label{eff}
\rho^{\rm (eff)} (r)
&=\frac{1}{4r^2}
\left\{
8(1-3B) B'\xi^{(1)}\psi'
+B
\left[
\psi'^2 r^2
-16(B-1) (\psi'^2\xi^{(2)}+\psi''\xi^{(1)})
\right]
\right\},
\nonumber\\
p_r^{\rm (eff)} (r)
&=\frac{B\psi'}{4}
\left[
\frac{8(3B-1)A'\xi^{(1)}}{r^2 A}
+\psi'
\right],
\nonumber\\
p_t^{\rm (eff)} (r)
&=\frac{B}{4rA^2}
\left\{
 A\psi'(12A'B'\xi^{(1)} -r A\psi')
+B
\left(
 -4A'^2 \xi^{(1)} \psi'
+8A \xi^{(1)} \psi' A''
+8AA'
 (\psi'^2\xi^{(2)}+\psi''\xi^{(1)})
\right)
\right\}.
\end{align}

\subsection{Boundary conditions, mass, and scalar charge}

The boundary conditions 
to integrate Eqs.~\eqref{eq_A} and \eqref{eq_psi} numerically
from (the vicinity of) the horizon 
can be fixed
by solving Eqs.~\eqref{eq_A} and \eqref{eq_psi}
in the vicinity of the horizon $r= r_h$:
\begin{align}
\label{bc}
A
&= 
a_1 (r-r_h)
+{\cal O}\left((r-r_h)^2\right),
\nonumber\\
B
&=
\frac{r_h \left(r_h^2-\sqrt{r_h^4 -96 \xi^{(1)} (\psi_0)^2}\right)}
      {48\xi^{(1)}(\psi_0)^2}
(r-r_h)
+{\cal O}\left((r-r_h)^2\right),
\nonumber\\
\psi
&=
  \psi_0
-\left(
 \frac{r_h^2-\sqrt{r_h^4 -96 \xi^{(1)}(\psi_0)^2}}
       {4r_h\xi^{(1)} (\psi_0)}
\right)
(r-r_h)
+{\cal O}\left((r-r_h)^2\right),
\end{align}
where 
$\psi_0:= \psi (r=r_h)$ is the amplitude of the scalar field at the horizon,
$a_1$ is an arbitrary constant,
and we choose the branch
that recovers the Schwarzschild solution in the limit of $\xi^{(1)} (\psi_0)\to 0$.
The condition for the existence of the scalarized BH solutions is given by 
\begin{align}
\label{exist}
r_h^4 >96 \xi^{(1)}(\psi_0)^2.
\end{align}
Here, without loss of generality we set
\begin{align}
a_1=
\frac{r_h \left(r_h^2-\sqrt{r_h^4 -96 \xi^{(1)} (\psi_0)^2}\right)}
      {48\xi^{(1)}(\psi_0)^2}.
\end{align}

After integrating Eqs.~\eqref{eq_A} and \eqref{eq_psi} to a sufficiently large radius,
$r\gg r_h$,
the solutions approach the asymptotic form 
which can be obtained by solving Eqs.~\eqref{eq_A} and \eqref{eq_psi}
in the limit of $r\to \infty$: 
\begin{align}
\label{bc2}
A&= 1-\frac{2M}{r}+\frac{M Q^2}{12r^3}
   +\frac{MQ}{6r^4}
\left(
MQ+24 \xi^{(1)}(\psi_\infty)
\right)
+{\cal O} \left(\frac{1}{r^5}\right),
\nonumber\\
B&=1-\frac{2M}{r}
+\frac{Q^2}{4r^2}+\frac{M Q^2}{4r^3}
 +\frac{MQ}{3r^4}
\left(
MQ+24 \xi^{(1)}(\psi_\infty)
\right)
+{\cal O} \left(\frac{1}{r^5}\right),
\nonumber\\
\psi&=
\psi_\infty
+\frac{Q}{r}+\frac{MQ}{r^2}
+\frac{1}{r^3}\left(\frac{4M^2Q}{3}-\frac{Q^3}{24}\right)
+\frac{M}{6r^4}
 \left(
 12 M^2 Q
-Q^3
-24M \xi^{(1)}(\psi_\infty)
 \right)
+{\cal O} \left(\frac{1}{r^5}\right),
\end{align}
where $\psi_\infty:=\psi (r\to \infty)$ is the asymptotic value of the scalar field,
and $M$ and $Q$ are the mass and the scalar charge, respectively.
From Eq.~\eqref{bc2}, $M$ and $Q$
can be evaluated as 
\begin{align}
\label{m_eva}
M&=\frac{r}{2}\left(1-B\right)\Big|_{r\to \infty},
\\
\label{q_eva}
Q&=-r^2 \psi'(r)\Big|_{r\to \infty}.
\end{align}
We note that 
because of the time reparametrization symmetry   
there is always the degree of freedom to rescale $A$
by some constant factor in Eqs. \eqref{bc} and \eqref{bc2}.
We emphasize that the physical quantities such as $M$ and $Q$
are not affected by the rescaling of $A$ at all.

For $\psi_0$ such that 
\begin{align}
\xi^{(1)} (\psi_0)=0,
\end{align}
the Schwarzschild solution $A(r)=B(r)=1-2M/r$ 
with a constant scalar field $\psi(r)=\psi_0$
is always a solution of the theory \eqref{esgb} with $Q=0$.

On the other hand,
for $\psi_0$ such that $\xi^{(1)}(\psi_0)\neq 0$,
$\psi(r)$ varies with $r$ and hence $Q$ does not vanish in general.
After solving Eqs.~\eqref{eq_A} and \eqref{eq_psi} for $A(r)$ and $\psi(r)$ iteratively
for sufficiently many different sets of the parameters,
$\psi_0$ and the coupling constants in $\xi(\phi)$,
we will find sets of them
that satisfy the boundary condition $\psi_\infty =0$ in the large distance limit.
For $\psi_0>0$,
from Eq. \eqref{bc}
the decreasing $\psi(r)$ in the vicinity of $r=r_h$
requires $\xi^{(1)} (\psi_0)>0$.
Similarly,
for $\psi_0<0$,
from Eq. \eqref{bc}
the increasing $\psi(r)$ in the vicinity of $r=r_h$
requires $\xi^{(1)} (\psi_0)<0$.
Assuming that $\xi(\phi)$ is an even function across $\phi=0$,
such as Eqs.~\eqref{quadratic} and \eqref{general},
without loss of generality we focus on the case of
\begin{align}
\label{firstorder}
\psi_0>0,
\qquad
\xi^{(1)} (\psi_0)>0.
\end{align}

\section{Scalarized black holes}
\label{sec3}

In this section,
we present our numerical scalarized BH solutions
for several choices of the coupling function $\xi(\phi)$.
In order to compare with the analysis in Sec. \ref{sec13},
we will consider 
the pure quadratic coupling~\eqref{quadratic}
and
the quartic order coupling~\eqref{general},
and investigate 
whether the inclusion of the gravitational backreaction 
modifies the properties of the solutions.
As mentioned in Sec. \ref{sec13},
the first model has been originally investigated in Refs. \cite{Silva:2017uqg,Blazquez-Salcedo:2018jnn},
and we will also confirm the consistency with them.

For the couplings \eqref{quadratic} and \eqref{general},
in order to find the scalarized BH solutions, 
we will vary the dimensionless parameters 
$(\eta/r_h^2,\psi_0)$
and 
$(\eta/r_h^2, \alpha, \psi_0)$,
respectively.
In our numerical analysis,
without loss of generality 
we will fix $r_h=1$ and integrate the equations \eqref{eq_A} and \eqref{eq_psi} 
from $r_s=r_h+10^{-5}$ with the boundary conditions \eqref{bc} to $r=5.0\times 10^4$. 
For such solutions, 
following Eqs.~\eqref{m_eva} and \eqref{q_eva},
$M$ and $Q$ are numerically evaluated at $r=5.0\times 10^4$.

\subsection{Quadratic coupling}
\label{sec31}

First, we review the case of the quadratic order coupling \eqref{quadratic}.
The condition \eqref{exist} is given by 
\begin{align}
\label{exist0}
r_h^4>6\eta^2\psi_0^2.
\end{align}
For the quadratic order coupling \eqref{quadratic},
the condition~\eqref{firstorder} is automatically satisfied.

\begin{figure}[h]
\unitlength=1.1mm
\begin{center}
  \includegraphics[height=6.0cm,angle=0]{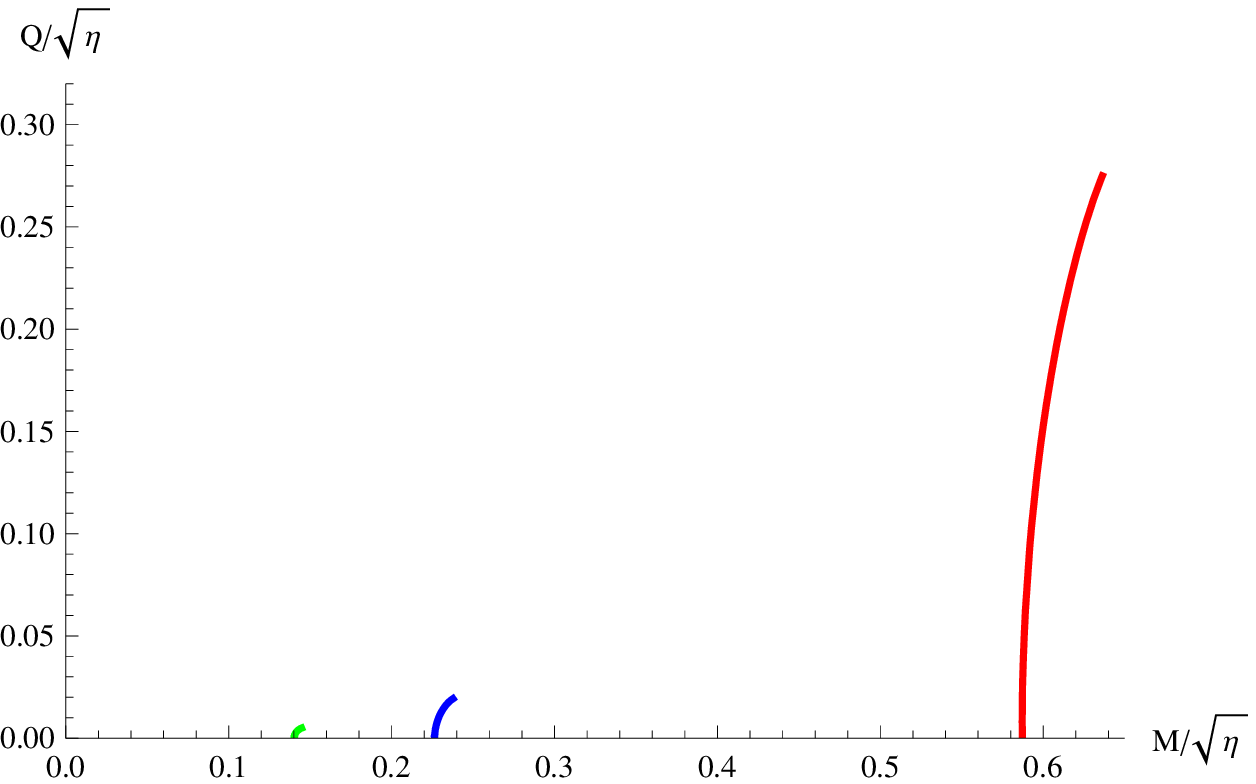}
  \includegraphics[height=4.0cm,angle=0]{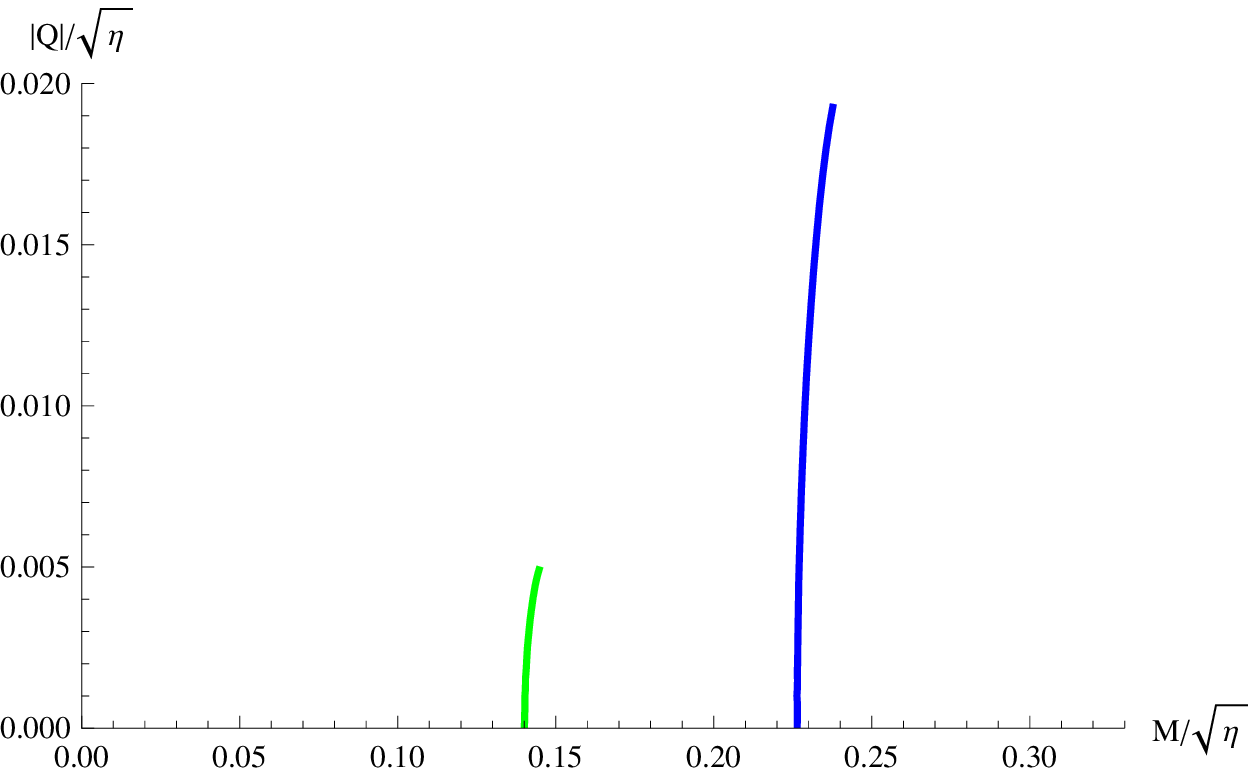}
\caption{
The absolute value of the scalar charge $|Q|$ divided by $\sqrt{\eta}$
is shown as the function of 
the mass $M$ divided by $\sqrt{\eta}$.
In the left larger panel,
the red, blue, and green curves correspond to  
the scalarized BH solutions
with zero, one, and two nodes
of the scalar field, respectively.
The right smaller panel 
corresponds to 
the enlarged display for the
scalarized BH solutions 
with one and two nodes.
We note that for $\psi_0>0$,
$Q>0$ for the scalarized BH solutions with an even number of nodes,
and $Q<0$ for those with an odd number of nodes.
For all the cases, 
as the value of $\psi_0$ increases, the value of $|Q|$ also increases.
}
  \label{figmq}
\end{center}
\end{figure}

In Fig.~\ref{figmq},
the absolute value of the scalar charge $|Q|$ divided by $\sqrt{\eta}$
is shown as the function of 
the mass $M$ divided by $\sqrt{\eta}$.
In the left larger panel,
the red, blue, and green curves correspond to  
the scalarized BH solutions
with zero, one, and two nodes
of the scalar field, respectively.
The right smaller panel 
corresponds to 
the enlarged display for the scalarized BH solutions with one and two nodes.
Although there are solutions with 
more than three nodes of the scalar field \cite{Blazquez-Salcedo:2018jnn},
we will not show them explicitly,
because of the similarity to the solutions with fewer numbers of nodes.
Because of the symmetry $\xi(-\phi)=\xi(\phi)$,
if $(g_{\mu\nu},\phi)$ is a solution 
then $(g_{\mu\nu},-\phi)$ is also a solution.
Thus, 
a scalarized BH solution with $(M, Q)$
is physically equivalent to that with $(M,-Q)$.
For $\psi_0>0$,
$Q>0$ for the scalarized BH solutions with an even number of nodes,
and $Q<0$ for those with an odd number of nodes.

From Fig.~\ref{figmq},
we find that
the branch of the scalarized BH solutions with zero nodes
bifurcates from the Schwarzschild solution, i.e., the axis of $Q=0$,
at $M/\sqrt{\eta}= 0.587$ ($\eta/M^2\approx 2.90$)
and
extends to $(M/\sqrt{\eta}, |Q|/\sqrt{\eta}) =(0.636, 0.275)$,
where the bound \eqref{exist0} is saturated.
The bifurcation point agrees with the case of the test field analysis.
For all the cases of scalarized BH solutions, 
as the value of $\psi_0$ increases,
the value of $|Q|$ also increases.
The difference because of the inclusion of the backreaction
is observed as the varying $M/\sqrt{\eta}$
and 
the existence of the upper bound on $|Q|$
resulting from the upper bound on $\psi_0$.
Similarly,
the scalarized BH solutions 
with one and two nodes
bifurcate from the Schwarzschild solutions
at $M/\sqrt{\eta}\approx 0.226$ ($\eta/M^2\approx 19.5$)
and 
at $M/\sqrt{\eta}\approx 0.140$ ($\eta/M^2\approx 50.9$),
and
extend to 
$(M/\sqrt{\eta}, |Q|/\sqrt{\eta})=(0.233, 0.0162)$
and 
$(M/\sqrt{\eta}, |Q|/\sqrt{\eta}) =(0.144, 0.00464)$,
respectively,
where the maximal charges are obtained
when Eq.~\eqref{exist0} is again saturated.
These results are consistent with those in Ref.~\cite{Silva:2017uqg,Blazquez-Salcedo:2018jnn}.

\begin{figure}[h]
\unitlength=1.0mm
\begin{center}
  \includegraphics[height=7.5cm,angle=0]{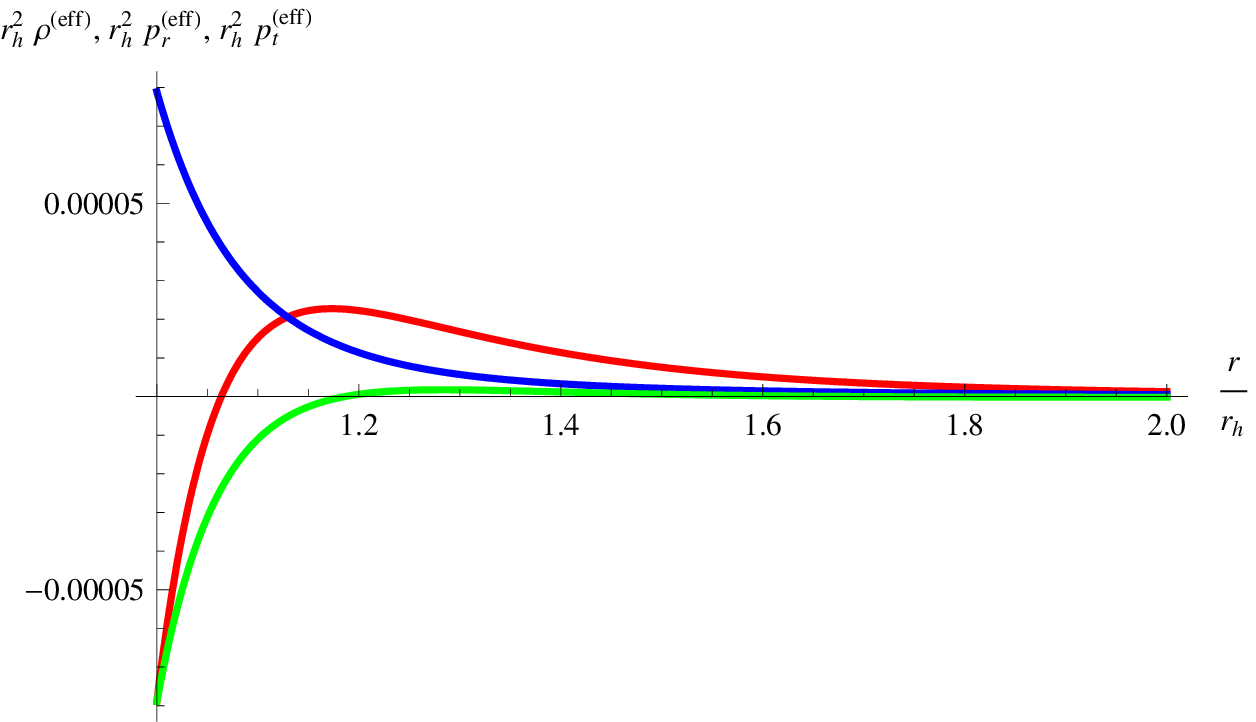}
  \includegraphics[height=4.5cm,angle=0]{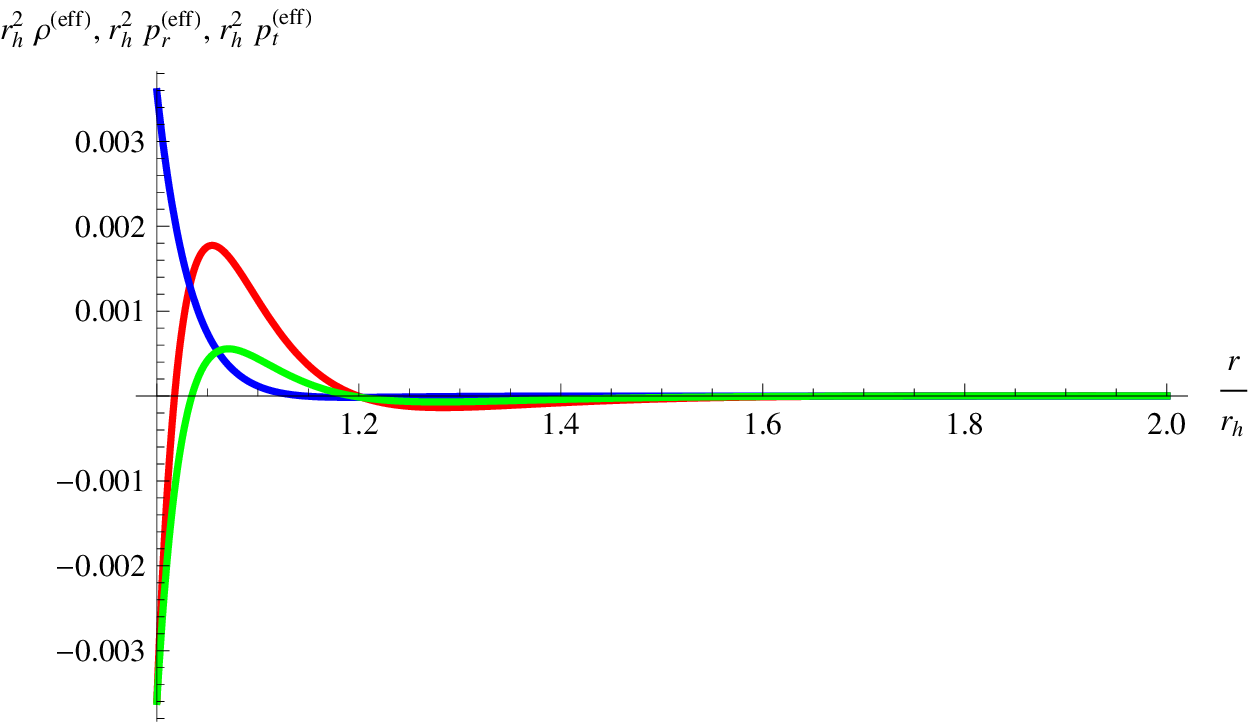}
  \includegraphics[height=4.5cm,angle=0]{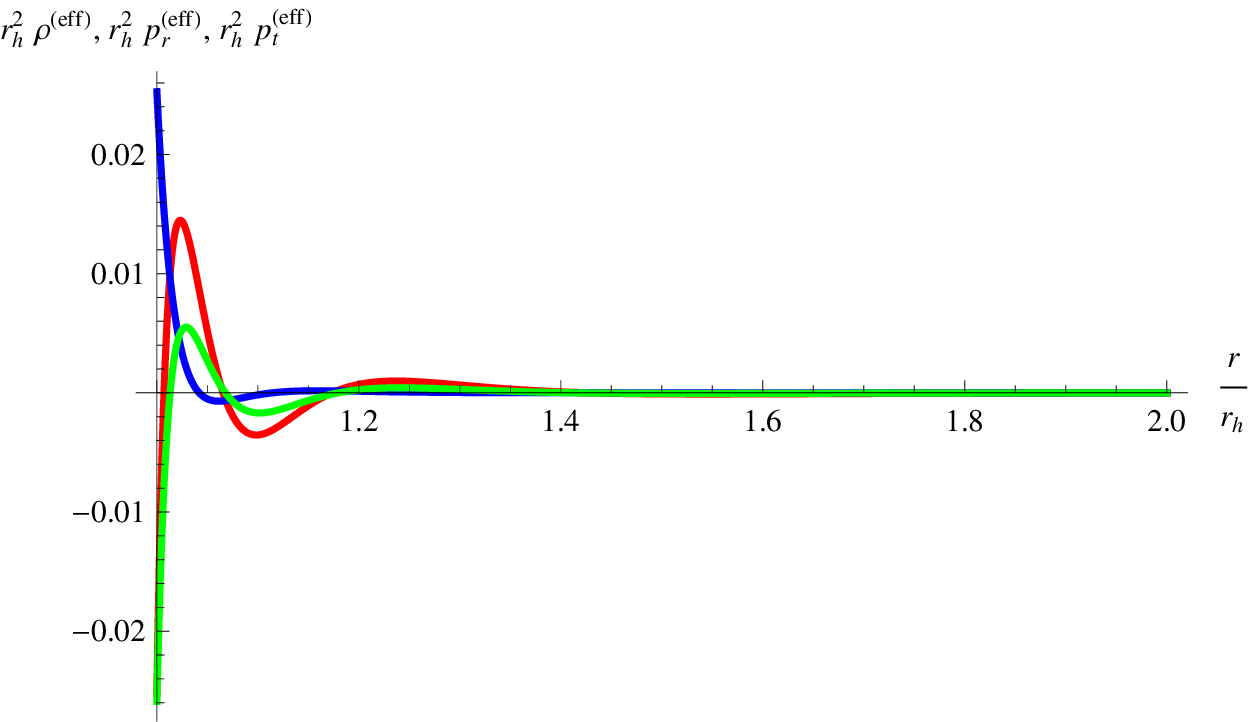}
\caption{
The effective energy density $\rho^{\rm (eff)} (r)$, 
the effective radial pressure $p_r^{\rm (eff)} (r)$,
and the effective tangential pressure $p_t^{\rm (eff)} (r)$ 
defined in Eq.~\eqref{eff},
which are multiplied by $r_h^2$,
are shown as the functions of $r/r_h$ for $\psi_0=0.01$.
The top panel 
corresponds to the cases of the scalarized BH solutions
with zero nodes of the scalar field, 
while the bottom-left and bottom-right panels
correspond to the cases of the scalarized BH solutions
with one and two nodes of the scalar field, respectively.
In each panel,
the red, blue, and green curves correspond 
to  
$\rho^{\rm (eff)} (r)$,
$p_r^{\rm (eff)} (r)$,
and 
$p_t^{\rm (eff)} (r)$,
respectively.
}
  \label{figem}
\end{center}
\end{figure}

In Fig.~\ref{figem},
the effective energy density $\rho^{\rm (eff)} (r)$, 
the effective radial pressure $p_r^{\rm (eff)} (r)$,
and the effective tangential pressure $p_t^{\rm (eff)} (r)$ 
defined in Eq.~\eqref{eff},
which are all multiplied by $r_h^2$,
are shown as the functions of $r/r_h$ for $\psi_0=0.01$.
The top panel 
corresponds to the cases of the scalarized BH solutions
with zero nodes of the scalar field, 
while the bottom-left and bottom-right panels
correspond to the cases of the scalarized BH solutions
with one and two nodes of the scalar field, respectively.
In each panel,
the red, blue, and green curves 
correspond to  
$\rho^{\rm (eff)}(r)$,
$p_r^{\rm (eff)}(r)$,
and 
$p_t^{\rm (eff)}(r)$,
respectively.
For all the solutions,
in the vicinity of the horizon $r=r_h$,
$\rho^{\rm (eff)}(r)<0$, 
$p_r^{\rm (eff)} (r)>0$, 
and
$p_t^{\rm (eff)}(r)<0$.
On the other hand,
irrespective of the number of nodes,
for a sufficiently large radius,
we always obtain 
$\rho^{\rm (eff)}(r)>0$ 
and  
$p_r^{\rm (eff)}(r)>0$,
while 
$p_t^{\rm (eff)}(r)<0$
after several oscillations.

In Ref.~\cite{Blazquez-Salcedo:2018jnn},
it was argued 
that all the Schwarzschild and scalarized BH solutions
in the scalar-tensor theories with the quadratic coupling \eqref{quadratic}
are unstable against the radial perturbation ($\ell=0$).
It was also shown
that
a scalarized BH solution with $n(\geq 0)$ nodes
possesses $(n+1)$ pure imaginary modes.
As we argued in Sec. \ref{sec132},
the existence of more pure imaginary modes
for the scalarized BHs with more nodes
can be understood in terms of the behavior of the effective potential
in the test field analysis \eqref{ep}.
For the scalarized BH solution with more nodes of the scalar field,
the effective potential for the radial perturbation,
Eq.~\eqref{ep} with $\ell=0$,
contains a deeper negative region in the vicinity of the horizon
for a larger positive value of $\eta/M^2$.
In Sec. \ref{sec133},
we speculated about
the way to construct stable scalarized BH solutions
by adding higher order powers of $\phi$ to the GB coupling function.
In Sec. \ref{sec32}, we will consider the general coupling function \eqref{general}.
We will argue the issue of the radial perturbation and stability 
in Sec. \ref{sec4}.

\subsection{General coupling}
\label{sec32}

We then consider the coupling \eqref{general}.
In the limit of $\alpha=0$,
we recover the scalarized BH solutions obtained in Sec. \ref{sec31}.
On the other hand, 
in the limit of $\alpha\gg 1$
we obtain the pure quartic order coupling \eqref{quartic}.
We will focus on the case of $\eta>0$,
but not specify the sign of $\alpha$.
From Eq.~\eqref{firstorder},
we obtain the bound
\begin{align}
\label{general2}
\alpha \psi_0^2>-\frac{1}{2},
\end{align}
which is the same as Eq.~\eqref{general2_test}
in the test field analysis in Sec. \ref{sec133}.
We note that 
for the coupling \eqref{general},
in addition to the Schwarzschild solution with $\psi=0$,
for $\alpha<0$
those with $\psi=\pm \sqrt{-1/(2\alpha)}$
are also the solutions of the theory \eqref{esgb},
as Eq.~\eqref{gr_sol} in the test field analysis.

\begin{figure}[h]
\unitlength=1.1mm
\begin{center}
  \includegraphics[height=5.0cm,angle=0]{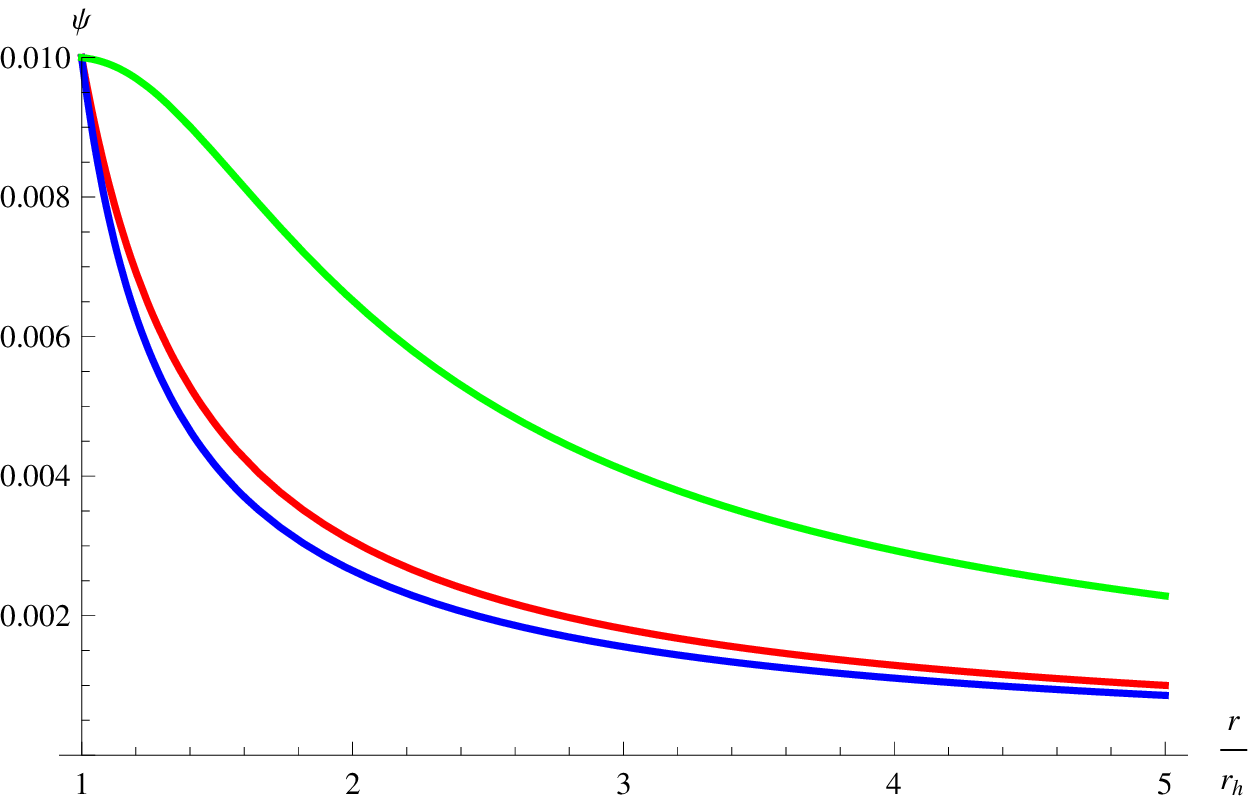}
  \includegraphics[height=5.0cm,angle=0]{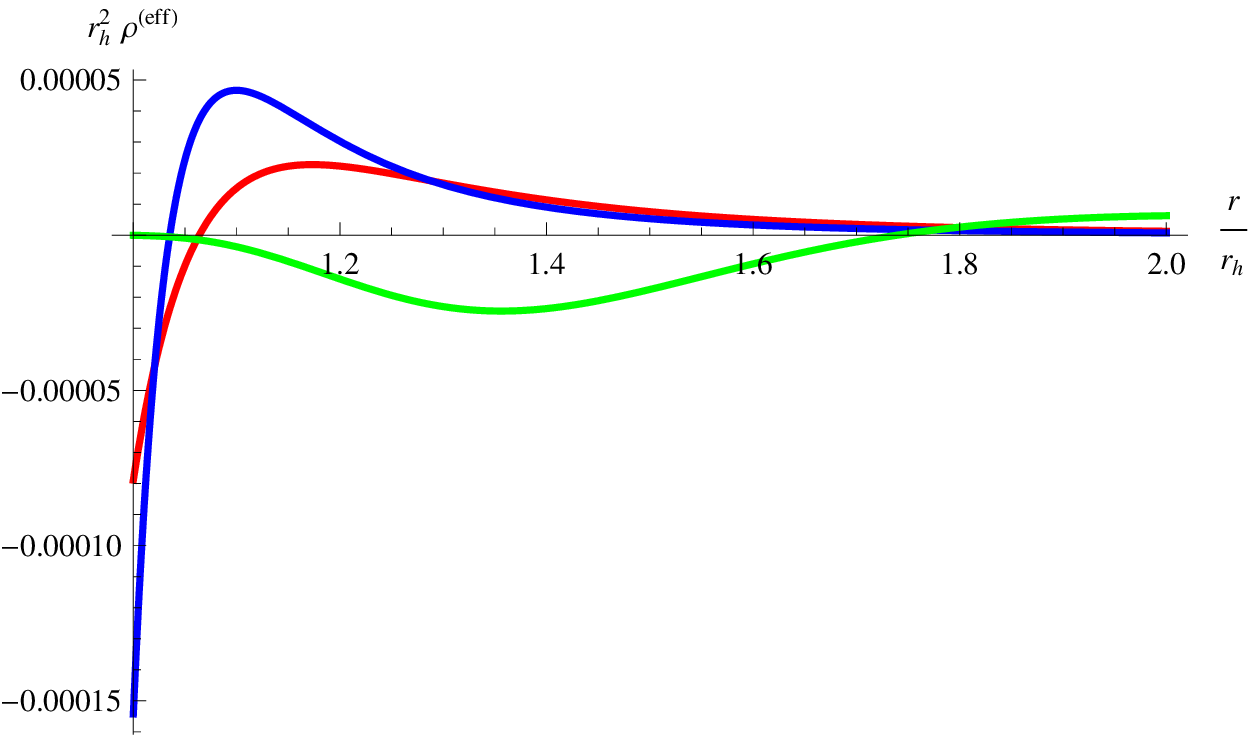}  
  \includegraphics[height=5.0cm,angle=0]{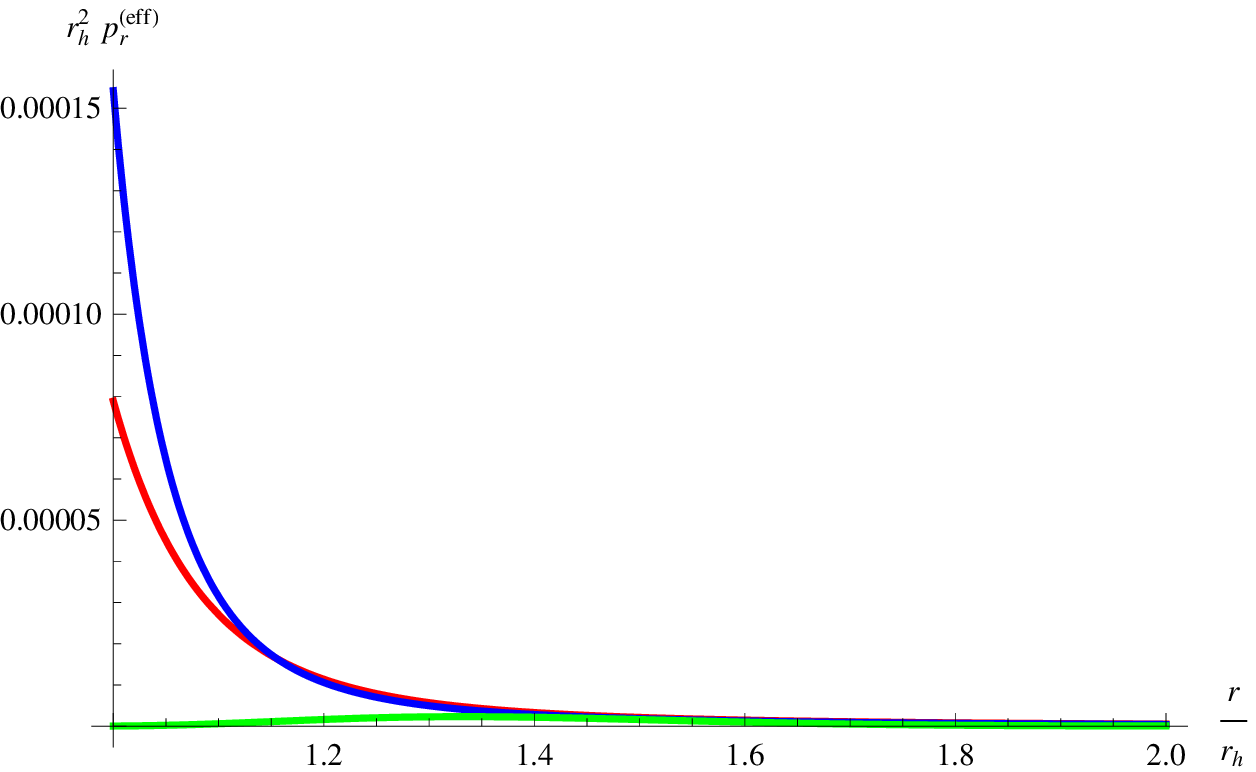}
  \includegraphics[height=5.0cm,angle=0]{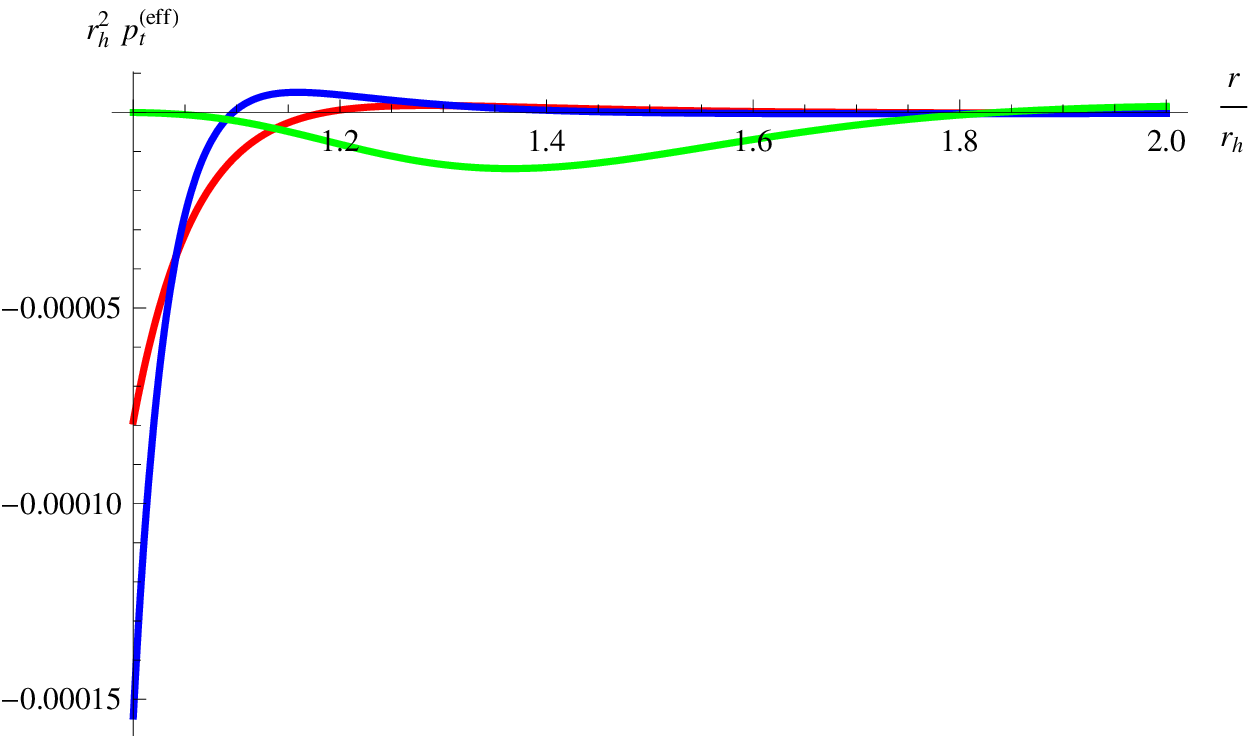}
\caption{
For the scalarized BHs with zero nodes of the scalar field,
the amplitude $\psi(r)$,
the effective energy density $\rho^{({\rm eff})}(r)$ multiplied by $r_h^2$,
the effective radial pressure $p^{({\rm eff})}_r(r)$ multiplied by $r_h^2$,
and 
the effective tangential pressure $p^{({\rm eff})}_t(r)$ multiplied by $r_h^2$
are shown as the functions of $r/r_h$
for $\psi_0=0.01$.
In each panel,
the red, blue, and green curves correspond
to the scalarized BH solutions
for 
$(\eta/r_h^2,\alpha)= (0.725, 0)$, $(0.338, 10000)$, $(7.31, -4990)$,
respectively.
}
  \label{figPsi}
\end{center}
\end{figure}

\begin{figure}[h]
\unitlength=1.1mm
\begin{center}
  \includegraphics[height=5.0cm,angle=0]{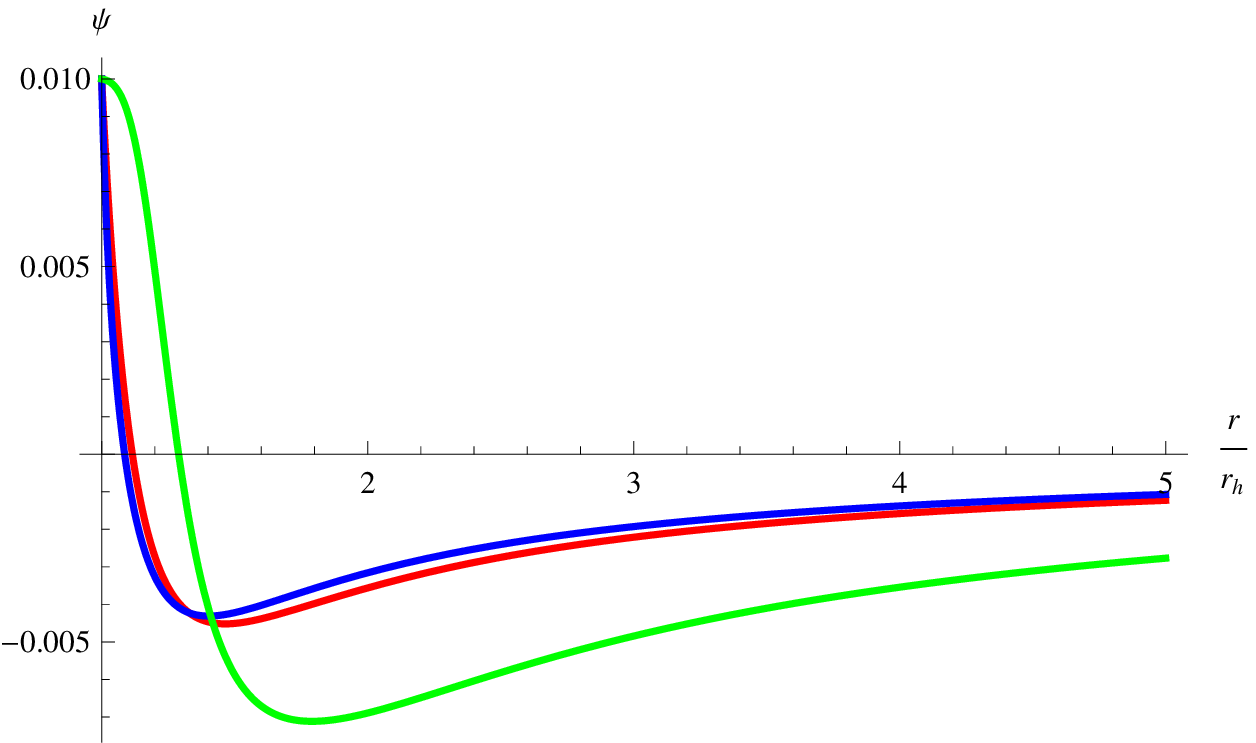}
  \includegraphics[height=5.0cm,angle=0]{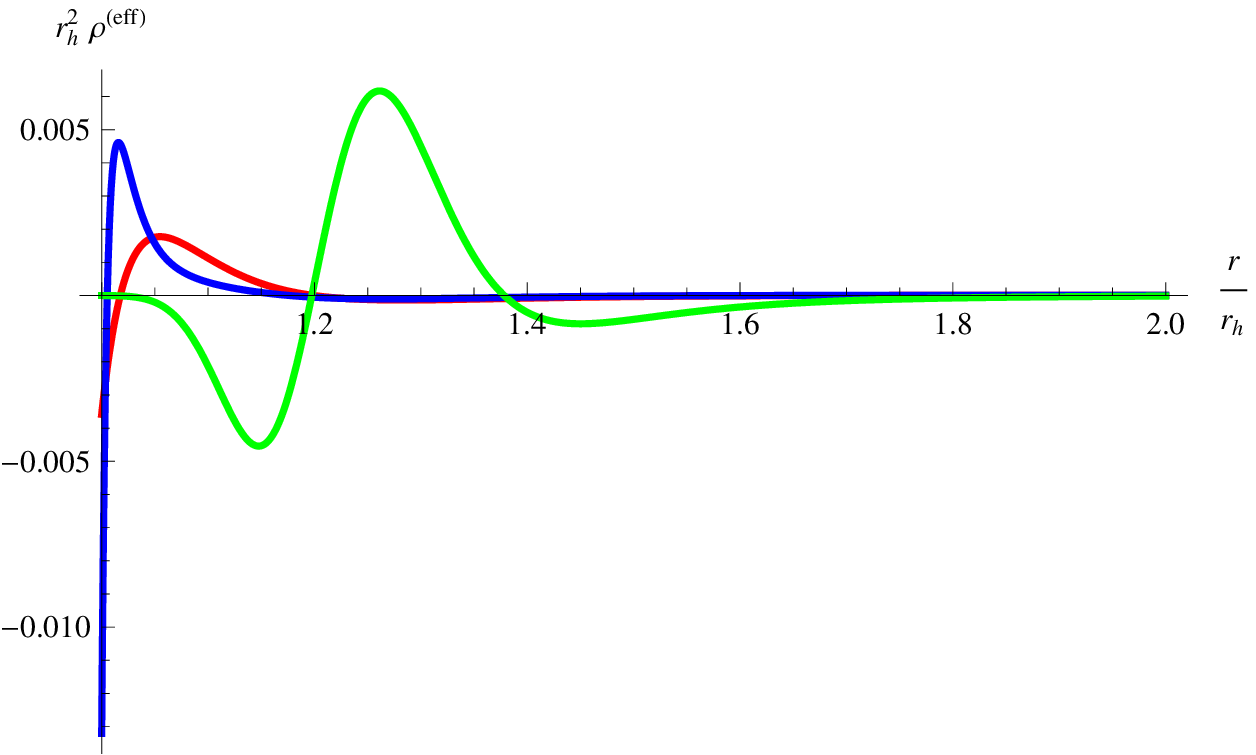}  
  \includegraphics[height=5.0cm,angle=0]{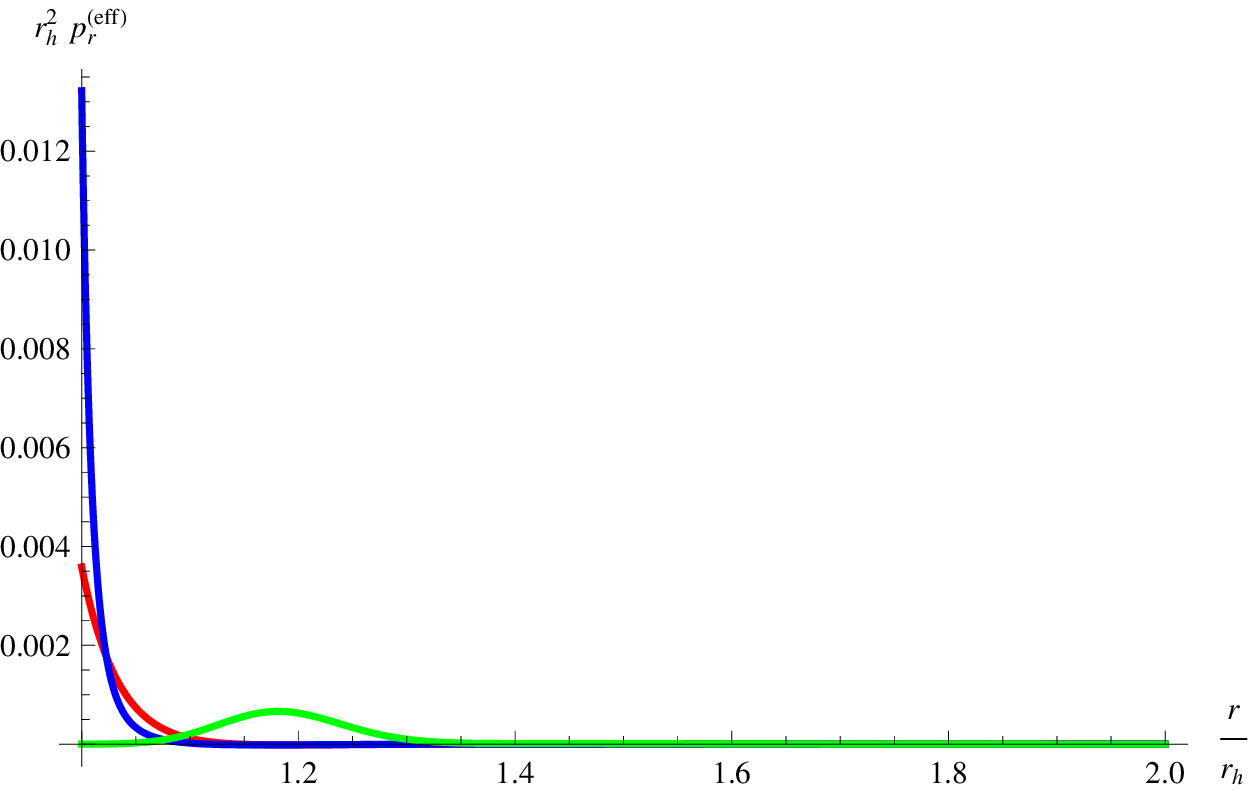}
  \includegraphics[height=5.0cm,angle=0]{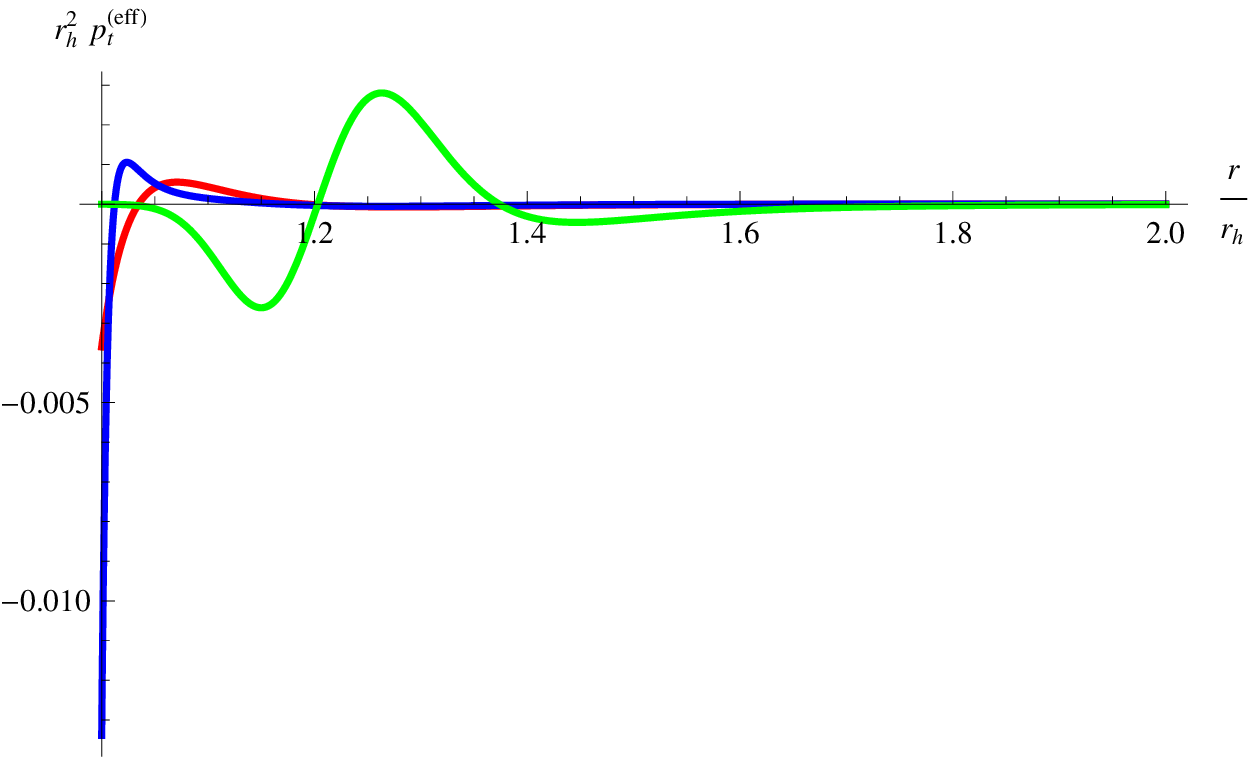}
\caption{
For the scalarized BHs with one node of the scalar field,
the amplitude $\psi(r)$,
the effective energy density $\rho^{({\rm eff})}(r)$ multiplied by $r_h^2$,
the effective radial pressure $p^{({\rm eff})}_r(r)$ multiplied by $r_h^2$,
and 
the effective tangential pressure $p^{({\rm eff})}_t(r)$ multiplied by $r_h^2$,
are shown as the functions of $r/r_h$
for $\psi_0=0.01$.
In each panel,
the red, blue, and green curves correspond
to the scalarized BH solutions
for 
$(\eta/r_h^2,\alpha)= (4.87, 0)$, $ (3.09, 10000)$, $(20.2, -4990)$,
respectively.
}
  \label{figPsi1}
\end{center}
\end{figure}

\begin{figure}[h]
\unitlength=1.1mm
\begin{center}
  \includegraphics[height=5.0cm,angle=0]{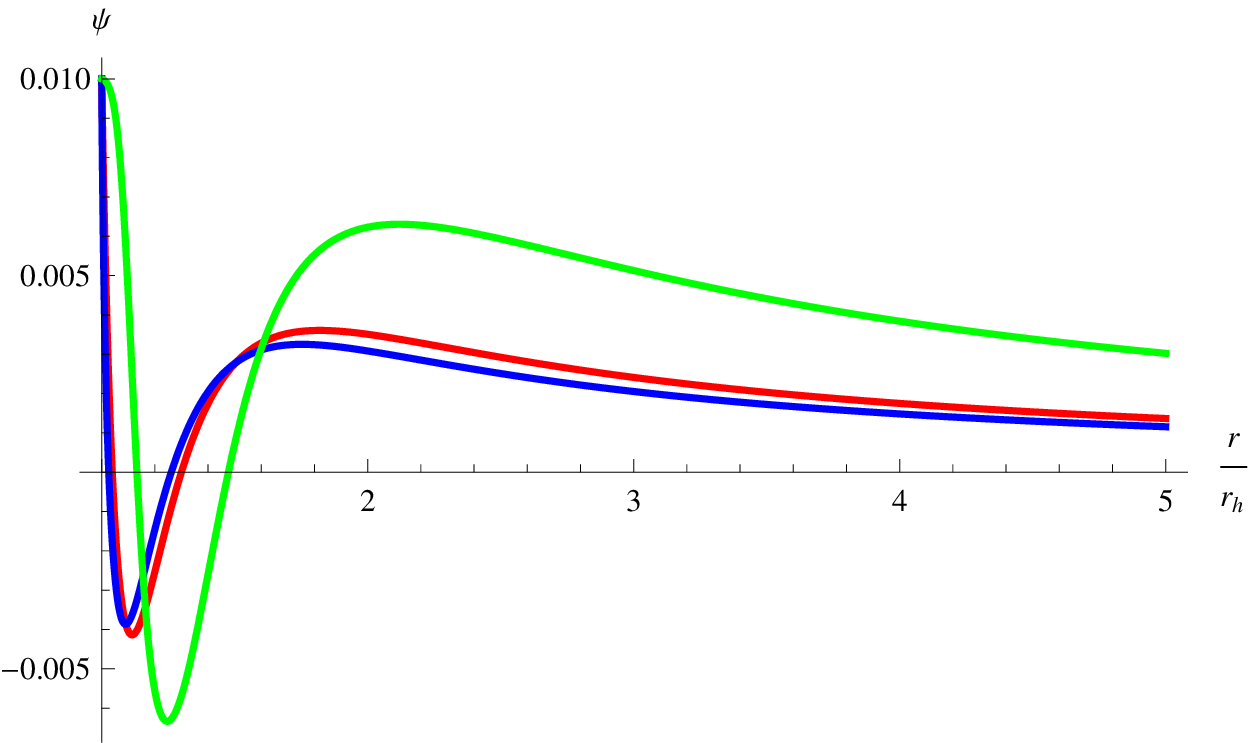}
  \includegraphics[height=5.0cm,angle=0]{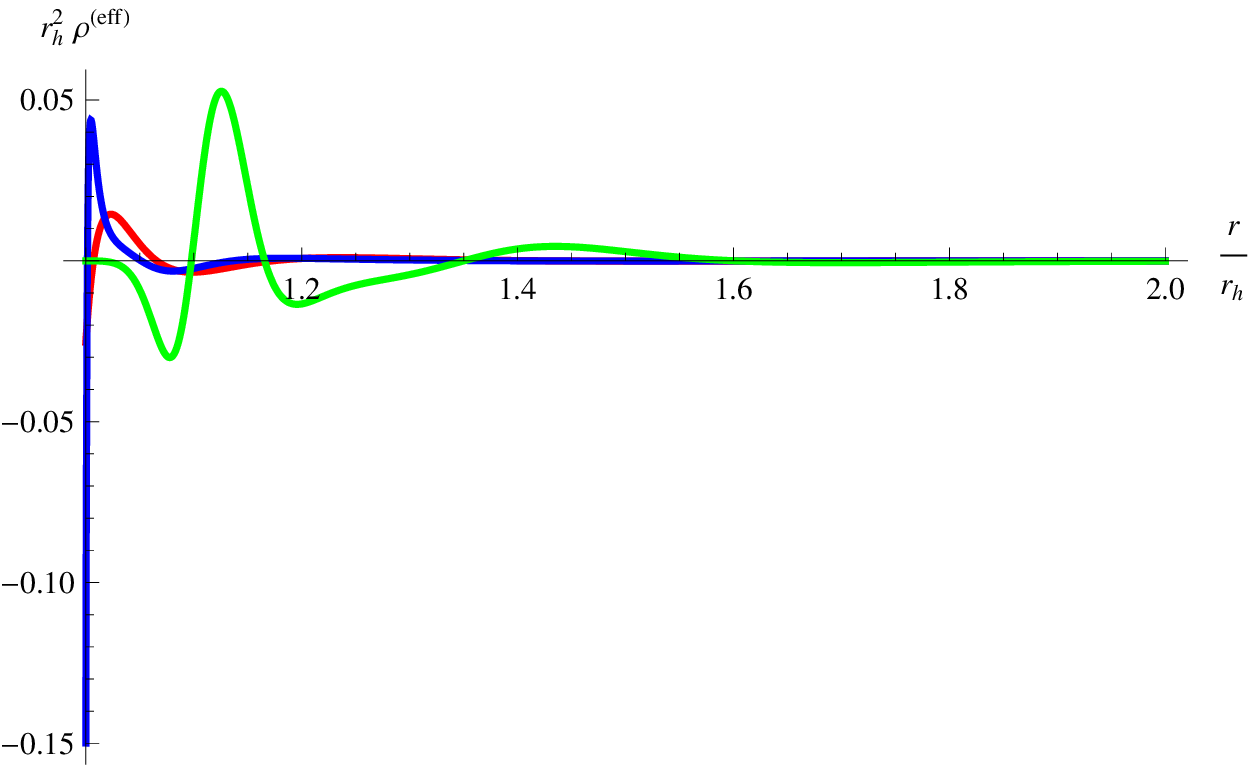}  
  \includegraphics[height=5.0cm,angle=0]{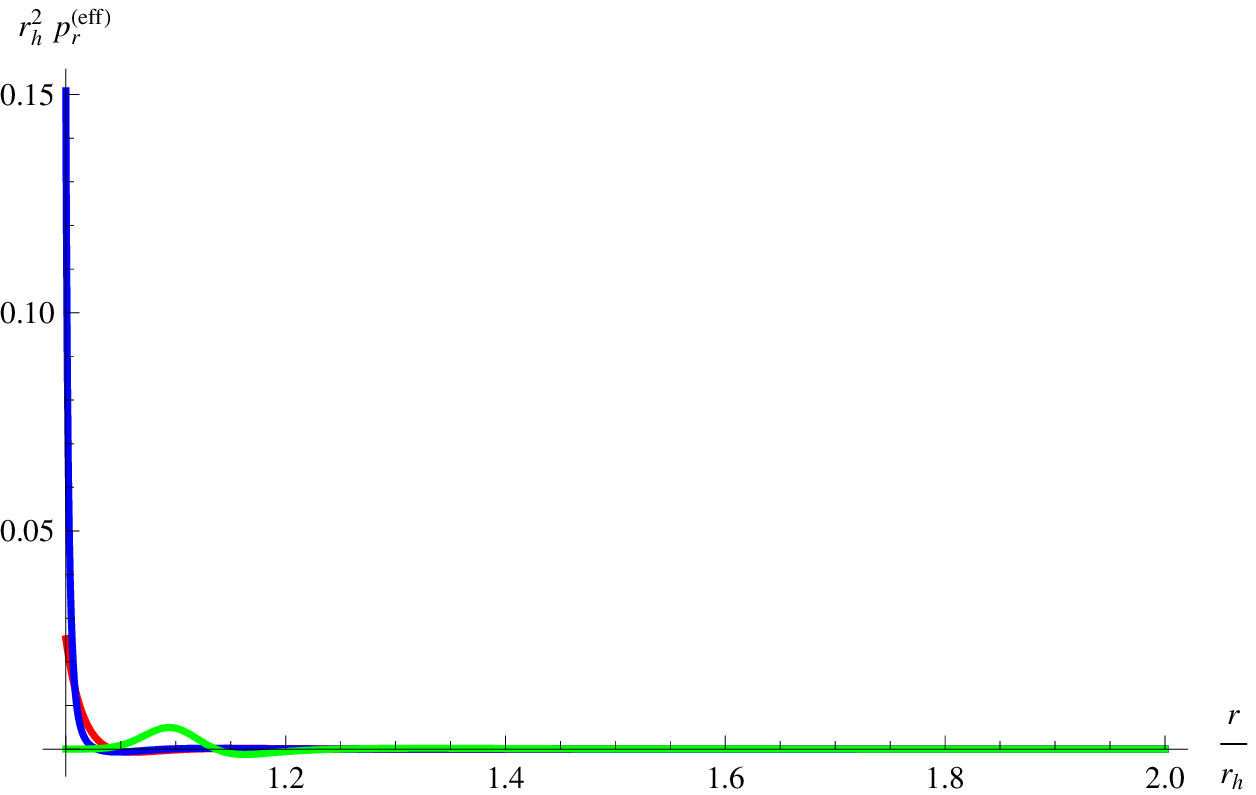}
  \includegraphics[height=5.0cm,angle=0]{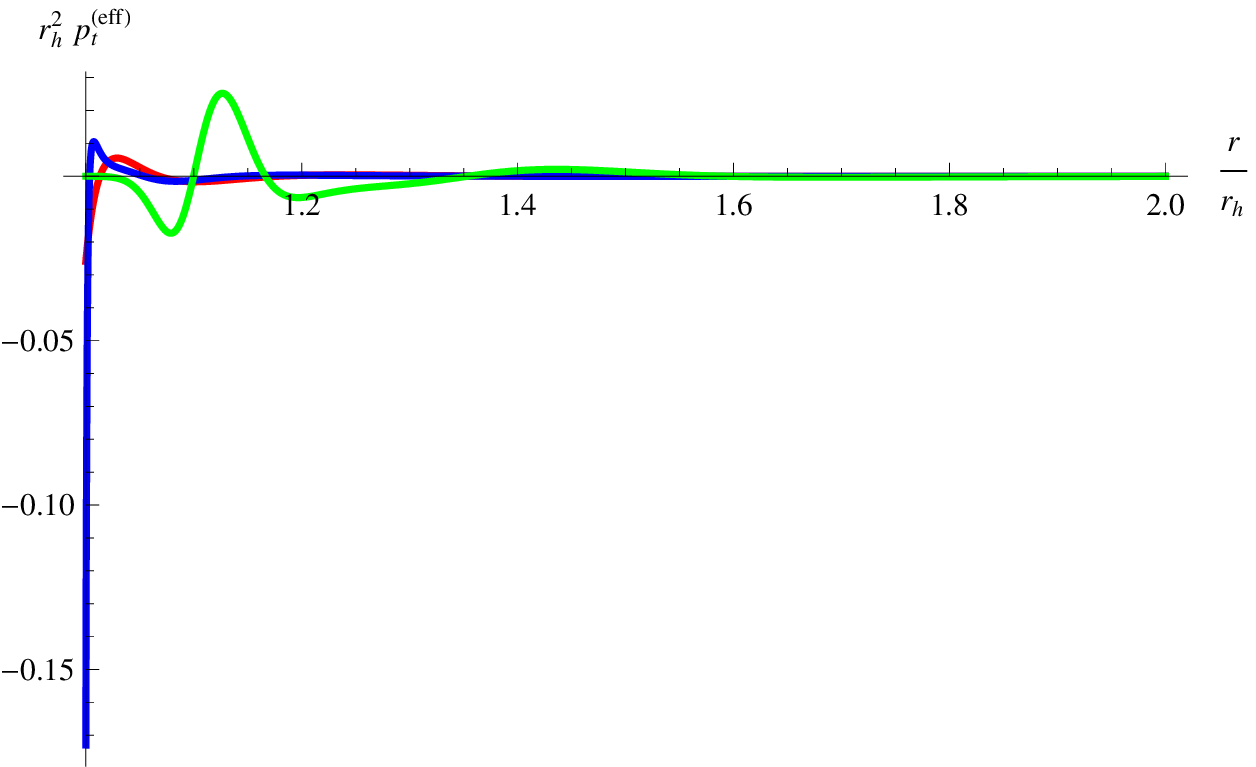}
\caption{
For the scalarized BHs with two nodes of the scalar field,
the amplitude $\psi(r)$,
the effective energy density $\rho^{({\rm eff})}(r)$ multiplied by $r_h^2$,
the effective radial pressure $p^{({\rm eff})}_r(r)$ multiplied by $r_h^2$,
and 
the effective tangential pressure $p^{({\rm eff})}_t(r)$ multiplied by $r_h^2$
are shown as the functions of $r/r_h$
for $\psi_0=0.01$.
In each panel,
the red, blue, and green curves correspond
to the scalarized BH solutions
for 
$(\eta/r_h^2,\alpha)= (12.7, 0)$, $(9.19, 10000)$, $(36.2, -4990)$,
respectively.
}
  \label{figPsi2}
\end{center}
\end{figure}

In Fig.~\ref{figPsi},
for the scalarized BHs with zero nodes of the scalar field,
the amplitude $\psi(r)$,
the effective energy density $\rho^{({\rm eff})}(r)$ multiplied by $r_h^2$,
the effective radial pressure $p^{({\rm eff})}_r(r)$ multiplied by $r_h^2$,
and 
the effective tangential pressure $p^{({\rm eff})}_t(r)$ multiplied by $r_h^2$
are shown as the functions of $r/r_h$
for $\psi_0=0.01$.
In each panel,
the red, blue, and green curves correspond
to the scalarized BH solutions
for 
$(\eta/r_h^2,\alpha)= (0.725, 0)$,  $(0.338, 10000)$, $(7.31, -4990)$,
respectively.
In the case of $\alpha<0$, 
$\psi(r)$ approaches zero more slowly than in the case of $\alpha=0$,
while
in the case of $\alpha>0$
faster than in the case of $\alpha=0$.
In the vicinity of the horizon, $1.1r_h\lesssim r\lesssim 1.7 r_h$, 
$\rho^{({\rm eff})}(r)$
becomes negative for $\alpha<0$,
while 
it becomes positive for $\alpha>0$.
$p^{({\rm eff})}_r(r)$
is positive both for $\alpha<0$ and for $\alpha>0$,
which is consistent with the analysis in Ref. \cite{Antoniou:2017acq}.
In the vicinity of the horizon, $1.1r_h\lesssim r\lesssim 1.8 r_h$, 
$p^{({\rm eff})}_t(r)$
becomes negative for $\alpha<0$.
In Figs.~\ref{figPsi1} and \ref{figPsi2},
the same plots for the scalarized BH solutions
with one and two nodes of the scalar field 
are shown, respectively.
Other than the oscillatory behaviors,
the essential behaviors 
of the scalarized BH solutions with one and two nodes of the scalar field
are the same as those with zero nodes of the scalar field.

\begin{figure}[h]
\unitlength=1.1mm
\begin{center}
  \includegraphics[height=7.5cm,angle=0]{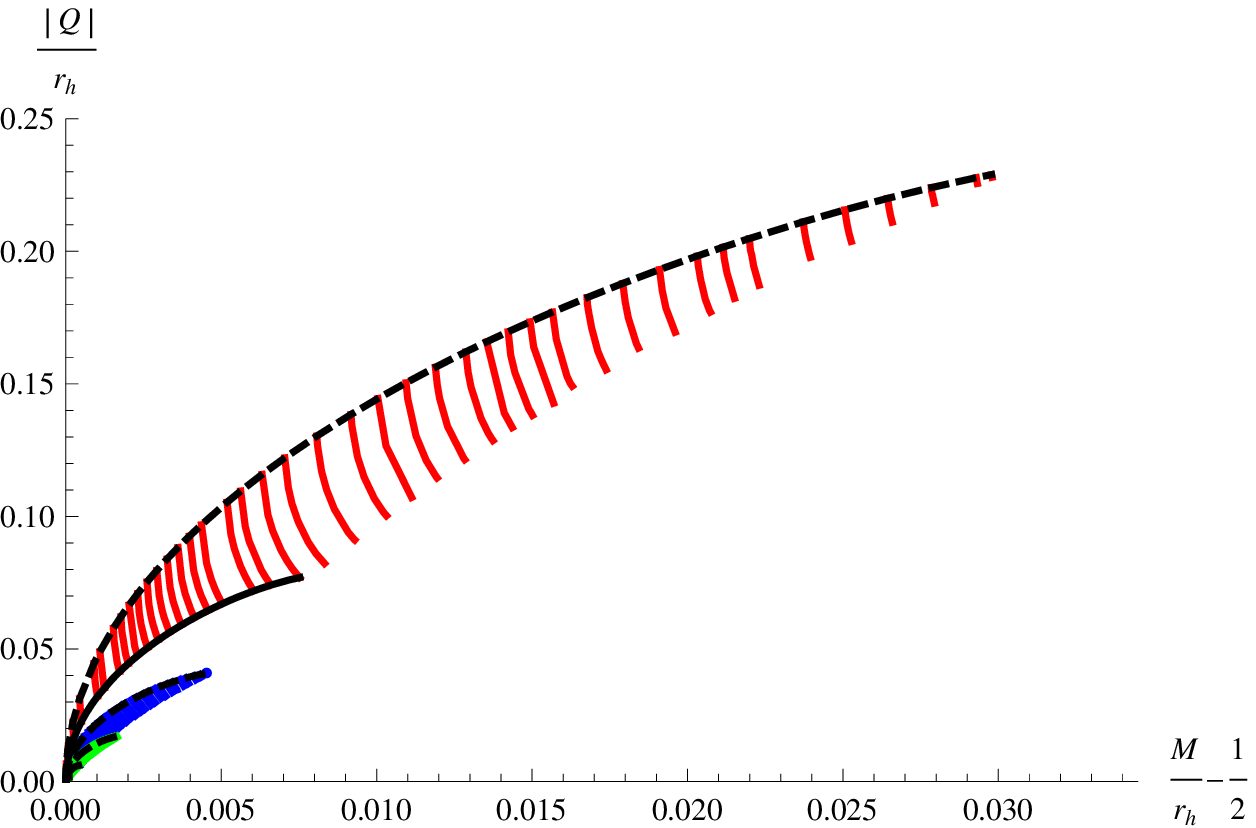}
  \includegraphics[height=5.0cm,angle=0]{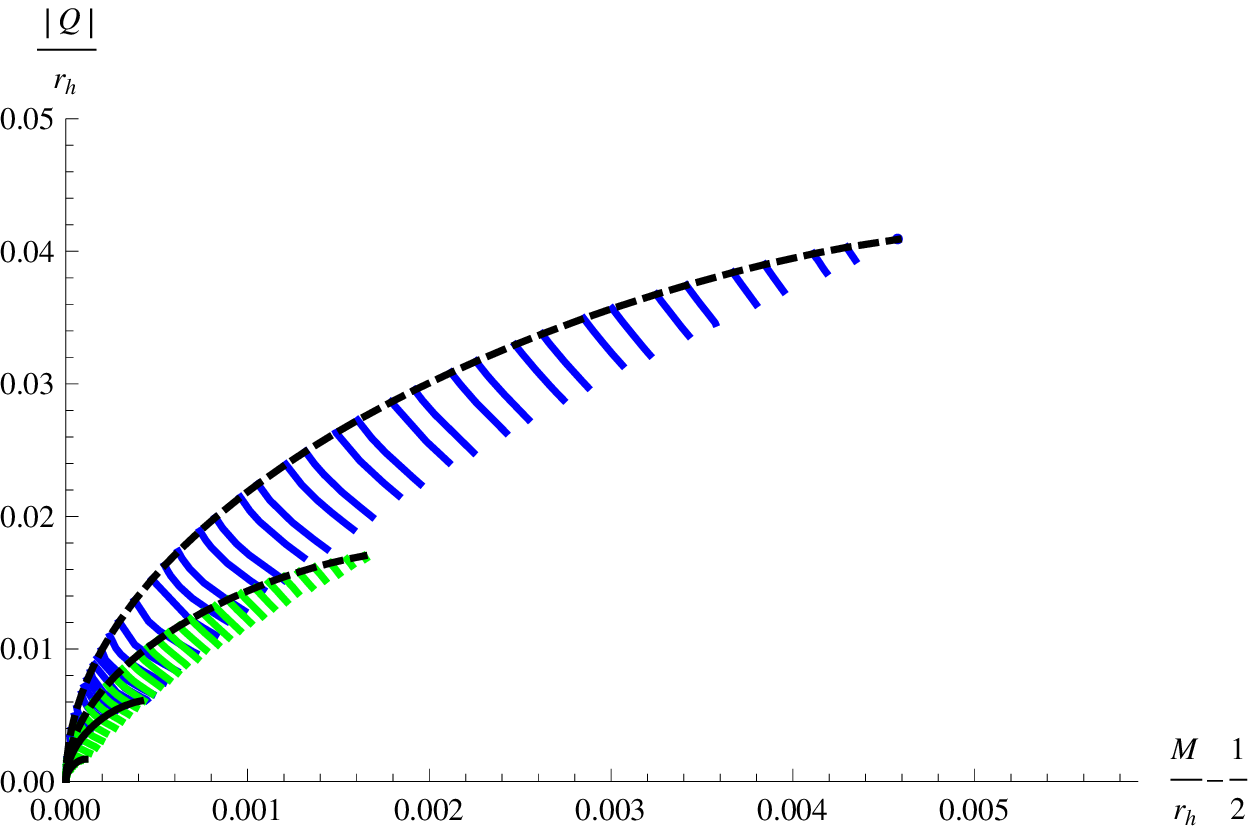}
  \includegraphics[height=5.0cm,angle=0]{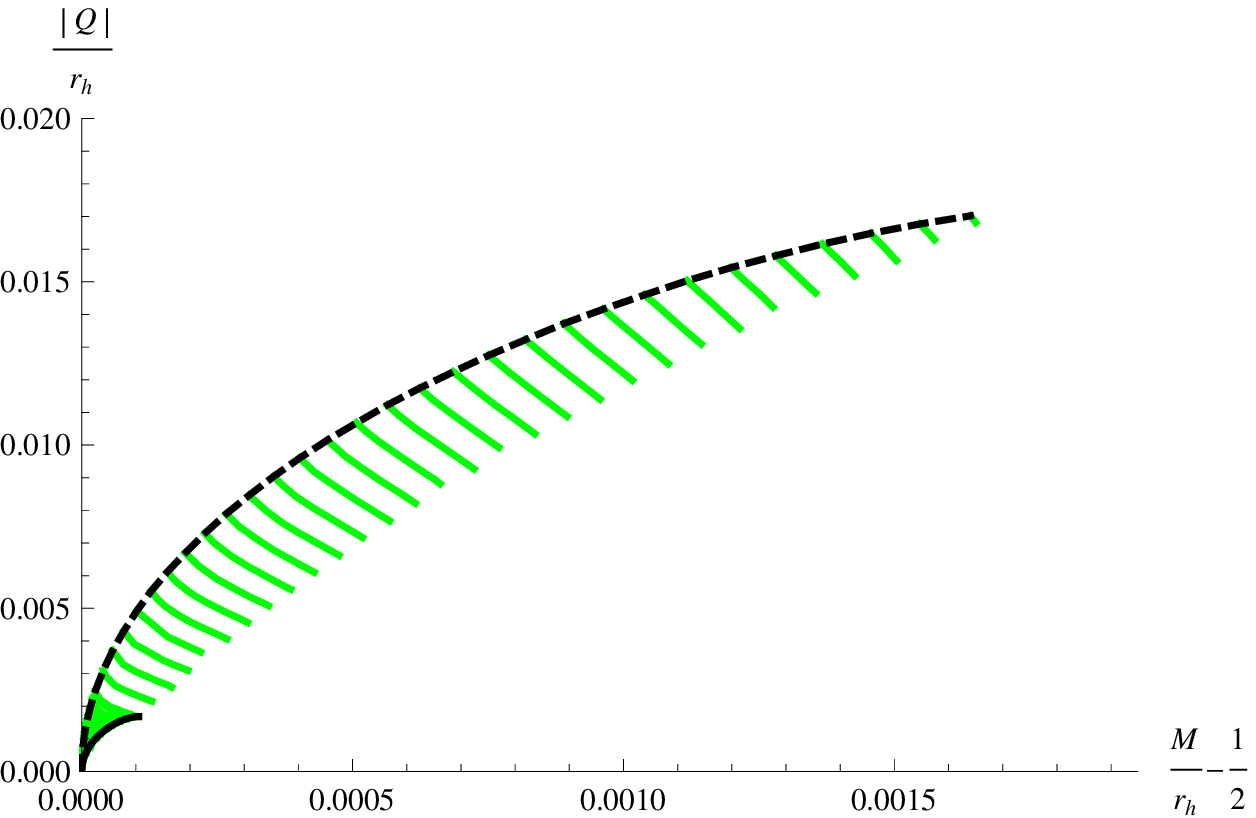}  
\caption{
The absolute value of the scalar charge $|Q|$ divided by $r_h$
is shown as the function of $(M/r_h-1/2)$.
The Schwarzschild solution without the scalar charge
corresponds to the origin $(0,0)$.
In the top panel,
the red, blue, and green curves
correspond to the scalarized BH solutions 
with zero, one, and two nodes of the scalar field,
respectively.
The bottom-left and bottom-right small panels
show the enlarged displays for the scalarized BH solutions
with one and two nodes of the scalar field,
respectively.
In each panel,
each curve represents the scalarized BH solutions
for the same value of $\psi_0$
and
the curves from the bottom
correspond
to those with the smaller values of $\psi_0$.
The scalarized BH solutions
in the pure quadratic coupling model ($\alpha=0$) 
correspond to the upper edges of the red, blue, and green regions,
respectively.
The bottom black solid curves in the red, blue, and green regions
correspond
to the scalarized BH solutions with the same number of nodes
for the pure quartic order coupling to the GB term ($\alpha\to \infty$) \eqref{quartic},
and 
the top black dashed curves 
correspond 
to those
for the pure quadratic order coupling to the GB term ($\alpha=0$) \eqref{quadratic}.
We note that for $\psi_0>0$,
$Q>0$ for the scalarized BH solutions with an even number of nodes,
and $Q<0$ for those with an odd number of nodes.
}
  \label{figmq2}
\end{center}
\end{figure}

In Fig.~\ref{figmq2},
for $\alpha\geq 0$
the absolute value of the scalar charge $|Q|$ divided by $r_h$
is shown as the function of $(M/r_h-1/2)$,
where the Schwarzschild solution without the scalar charge
corresponds to the origin $(0,0)$.
In the top panel,
the red, blue, and green curves
correspond to the scalarized BH solutions 
with zero, one, and two nodes of the scalar field,
respectively.
The bottom-left and bottom-right panels
show the enlarged displays for the scalarized BH solutions
with one and two nodes of the scalar field,
respectively.
In each panel,
each curve represents the scalarized BH solutions
for the same value of $\psi_0$
and
the curves from the bottom
correspond
to those with the smaller values of $\psi_0$.
The scalarized BH solutions
in the pure quadratic coupling model ($\alpha=0$) 
correspond to the upper edges of the red, blue, and green regions,
respectively.
The bottom black solid curves in the red, blue, and green regions
correspond
to the scalarized BH solutions with the same number of nodes
for the pure quartic order coupling to the GB term ($\alpha\to \infty$) \eqref{quartic},
and 
the top black dashed curves 
correspond 
to those for the pure quadratic order coupling to the GB term ($\alpha=0$) \eqref{quadratic}.
In the unit of $r_h$,
for a fixed value of $\psi_0$
as $\alpha$ increases positively,
$M/r_h$ increases while $|Q|/r_h$ decreases.

In Fig.~\ref{figmq2},
we find that 
the black curve for each value for the pure quartic order coupling \eqref{quartic}
covers only a part of the lower edge
of the corresponding region for the coupling \eqref{general}.
This indicates 
that for a larger fixed value of $\psi_0$
the scalarized BH solutions exist
only up to a finite value of $\alpha>0$.
For a value of $\psi_0$ below some critical value
the scalarized BH solutions exist for an arbitrarily large value of $\alpha$,
and hence the limit to the pure quartic order coupling model \eqref{quartic} exists.
The existence of such a critical value of $\psi_0$
can be understood in terms of 
the condition for the existence of the solution \eqref{exist}.
For the coupling function \eqref{general},
the condition Eq.~\eqref{exist} reduces to 
\begin{align}
\label{exist2}
r_h^4>6\eta^2\psi_0^2\left(1+2\alpha\psi_0^2\right)^2.
\end{align}
From Eq. \eqref{general2} with $\psi_0>0$,
in the regime where the quartic order term dominates the quadratic one
the bound \eqref{exist2} reduces to
\begin{align}
\eta<\frac{r_h^{2}}
             {2\sqrt{6}\alpha\psi_0^3}.
\end{align}
Assuming that the limit to the case of the pure quartic order coupling with
the coupling constant $\lambda_0$ exists
in the large $\alpha$ limit,
$\eta$ and $\alpha$ giving a scalarized BH solution 
are related by a scaling relation $\eta = \lambda_0/\alpha$.
Hence, 
the existence of the limit to the pure quartic order coupling requires
$r_h^{2}/ (2\sqrt{6}\alpha\psi_0^3) >\lambda_0/\alpha$,
which gives rise to the critical value of $\psi_0$:
\begin{align}
\label{psi_bound}
\psi_0<\left(\frac{r_h^2}{2\sqrt{6}\lambda_0}\right)^{\frac{1}{3}}.
\end{align}

\begin{figure}[h]
\unitlength=1.1mm
\begin{center}
  \includegraphics[height=7.5cm,angle=0]{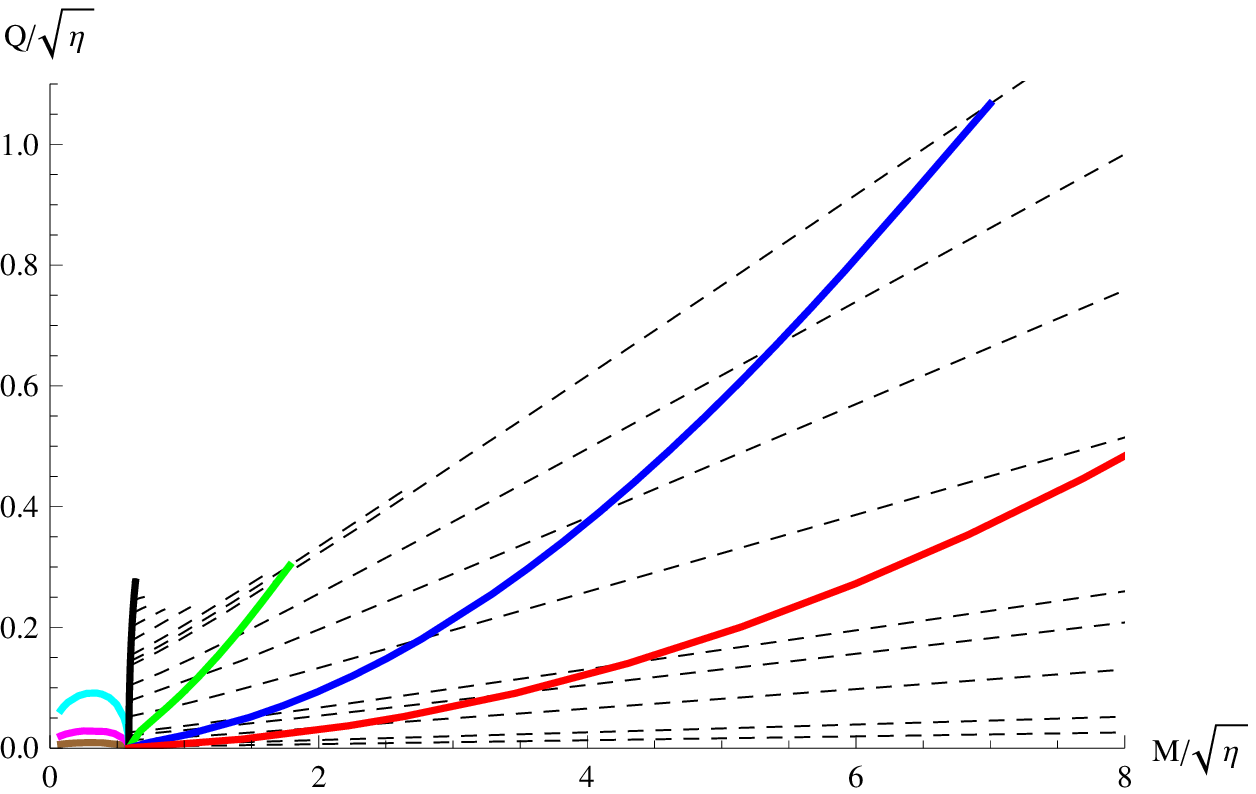}
\caption{
The scalar charge $Q$ divided by $\sqrt{\eta}$
for scalarized BH solutions with zero nodes of the scalar field
is shown as the function of 
the mass $M$ divided by $\sqrt{\eta}$
for fixed values of $\alpha$.
The red, blue, green, black, cyan, magenta, and brown curves
correspond 
to the cases of 
$\alpha=20000, 2000, 100, 0, -10, -100, -1000$,
respectively.
The dashed curves represent
scalarized BH solutions for each constant value of $\psi_0$
for $\alpha>0$,
where the upper curves correspond to the larger values of $\psi_0$.
For $\alpha<0$,
we have chosen the maximal value of $\psi_0$ to
$\sqrt{-1/(2\alpha)} (1- 10^{-8})$.}
  \label{figMQalpha}
\end{center}
\end{figure}

In Fig.~\ref{figMQalpha},
the scalar charge $Q$ divided by $\sqrt{\eta}$
for scalarized BH solutions with zero nodes of the scalar field
is shown as the function of 
the mass $M$ divided by $\sqrt{\eta}$
for fixed values of $\alpha$.
The red, blue, green, black, cyan, magenta, and brown curves
correspond 
to the cases of 
$\alpha=20000, 2000, 100, 0, -10, -100, -1000$,
respectively.
The dashed curves represent
scalarized BH solutions for each constant value of $\psi_0$
for $\alpha>0$,
where the upper curves correspond to the larger values of $\psi_0$.
For all values of $\alpha$,
the branches of the scalarized BH solutions with zero nodes
bifurcate from the Schwarzschild solution $Q=0$,
at $M/\sqrt{\eta}= 0.587$ for $\psi_0=0$,
and as $\psi_0$ increases,
the scalar charge $|Q|/\sqrt{\eta}$ increases.
This confirms our argument in Sec. \ref{sec131}
that the onset of the tachyonic instability of the Schwarzschild solution
is determined by the quadratic order term in $\xi(\phi)$.
For $\alpha<0$,
we have chosen the maximal value of $\psi_0$ to
$\sqrt{-1/(2\alpha)} (1- 10^{-8})$.

\begin{figure}[h]
\unitlength=1.1mm
\begin{center}
  \includegraphics[height=7.5cm,angle=0]{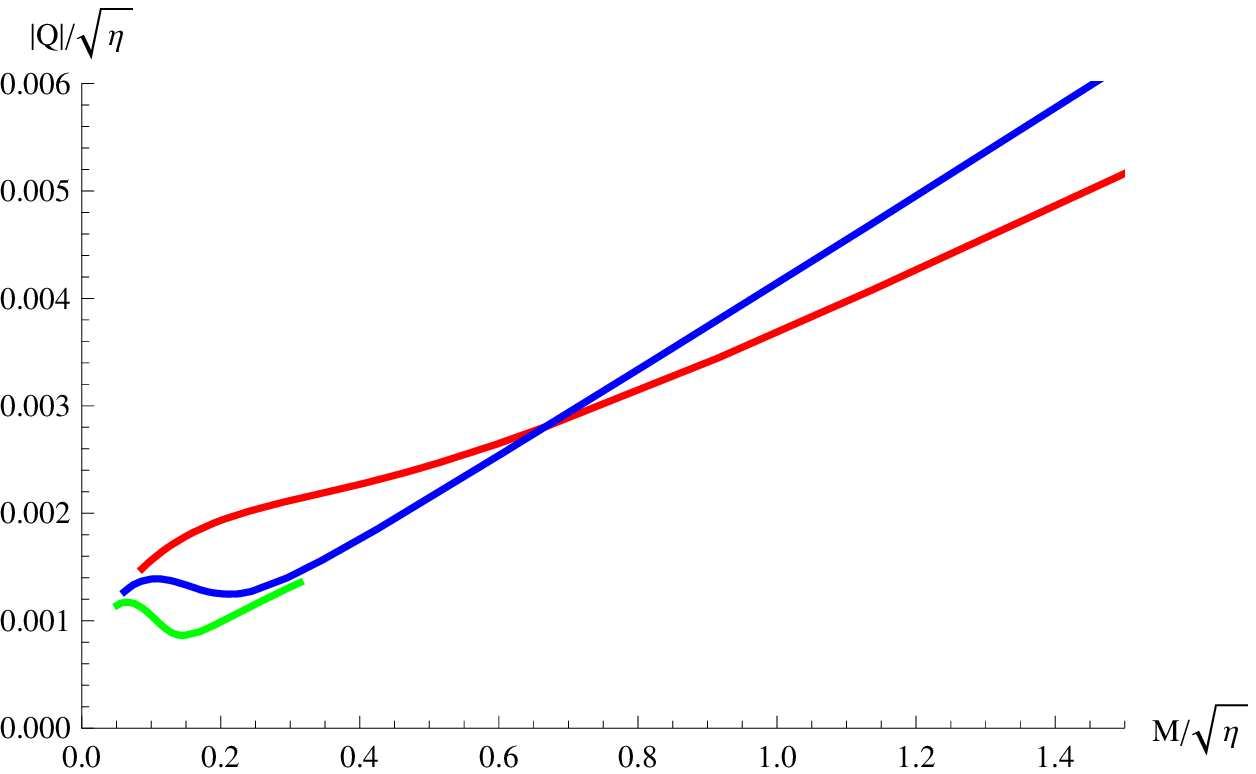}
\caption{
The absolute value of the scalar charge $|Q|$ divided by $\sqrt{\eta}$
is shown as the function of the mass $M$ divided by $\sqrt{\eta}$
for $\psi_0=0.005$.
The red, blue  and green curves
correspond to 
the solutions of the scalar field 
with zero, one, and two nodes,
respectively.
In the limit of $\alpha\to \infty$,
the red and blue curves for the scalarized BH solutions with zero and one node
approach 
the straight lines
corresponding to the scalarized BH solutions with the same number of nodes
for the pure quartic coupling \eqref{quartic}.
For $\alpha<0$,
we have chosen the minimal value of $\alpha $ to $(1/\psi_0^2) (-0.5 + 10^{-7})$.
}
  \label{figmqgen}
\end{center}
\end{figure}

In Fig.~\ref{figmqgen},
the absolute value of the scalar charge $|Q|$ divided by $\sqrt{\eta}$
is shown as the function of the mass $M$ divided by $\sqrt{\eta}$
for $\psi_0=0.005$,
for which all the scalarized BH solutions with zero, one, and two nodes of the scalar field exist.
The red, blue, and green curves
correspond to the scalarized BH solutions 
with zero, one, and two nodes of the scalar field, respectively.
In the limit of $\alpha\to \infty$,
the red and blue curves for the scalarized BH solutions with zero and one node
approach the straight lines
corresponding to the scalarized BH solutions with the same number of nodes
for the pure quartic coupling \eqref{quartic}.
We find that the relation between $M$ and $Q$
is very similar to Fig.~\ref{figtestgen} in Sec. \ref{sec13} for the test field analysis,
except for the existence of the upper bound on $\alpha$
for the scalarized BH solutions with two nodes.
We expect that 
the reason why the full analysis is similar to the test field analysis is
because of the existence of the condition \eqref{exist2},
which does not allow a large value of $\psi_0$
for which the backreaction 
would spoil the essential features obtained from the analysis in Sec.~\ref{sec13}.

\begin{figure}[h]
\unitlength=1.1mm
\begin{center}
  \includegraphics[height=7.5cm,angle=0]{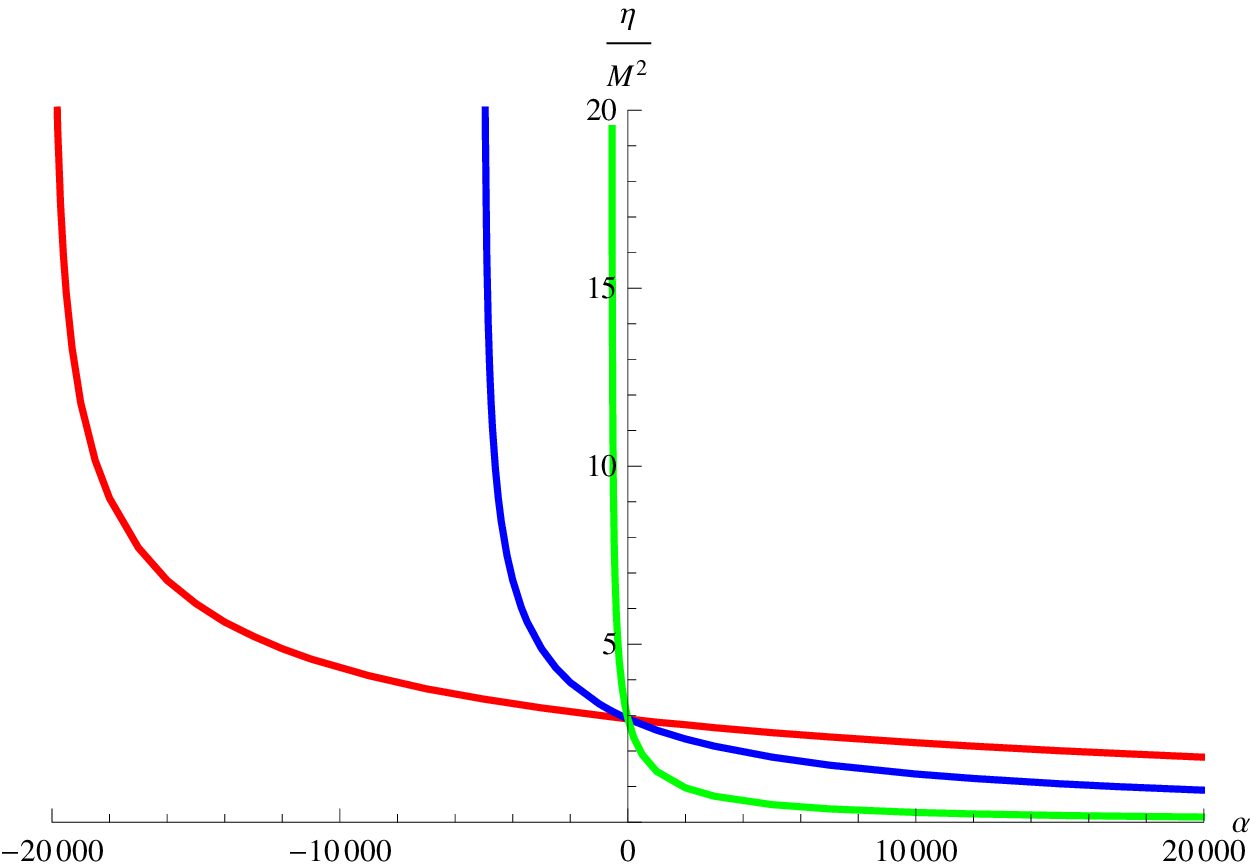}
\caption{For scalarized BHs with zero nodes of the scalar field,
$\eta/M^2$ is shown as the function of $\alpha$
for several values of $\psi_0={\cal O} (0.01)$.
The red, blue, and green curves
correspond to 
the cases of $\psi_0=0.005$, $0.01$, and $0.03$,
respectively.
}
  \label{figetaalpha}
\end{center}
\end{figure}

In Fig.~\ref{figetaalpha},
for scalarized BHs with zero nodes of the scalar field,
$\eta/M^2$ is shown as the function of $\alpha$
for several values of $\psi_0={\cal O}(0.01)$.
The red, blue, and green curves
correspond to 
the cases of $\psi_0=0.005$, $0.01$, and $0.03$,
respectively.
For a given $\psi_0$,
as $\alpha$ increases, 
$\eta/M^2$ decreases.
The value of $\eta/M^2$ blows up at $\alpha=-1/(2\psi_0^2)$,
where the bound \eqref{general2} is saturated.
We note that 
for a solar-mass BH with $M= {\cal O}(GM_\odot /c^2)={\cal O} (1{\rm km})$,
$\eta/M^2>1$ if $\sqrt{\eta^{-1}}<10^{-11}{\rm eV}$,
and 
for a BH with $M={\cal O} ((G/c^2)10^{15} {\rm g})={\cal O} (10^{-13} {\rm cm})$,
which is considered as the lowest mass of primordial BHs \cite{Carr:2016drx},
$\eta/M^2>1$ if $\sqrt{\eta^{-1}}<10^8 {\rm eV}$,
for which
the scalarized BHs can be realized also for $\alpha<0$.

In Sec. \ref{sec4},
we will analyze the radial perturbation of the scalarized BH solutions obtained in this section.
We will find that 
the similarity of the full analysis to the test field one
will also hold at the level of the linear perturbation,
and 
scalarized BHs which are radially stable 
exist only for the case of $\alpha<0$.

\section{Radial perturbation}
\label{sec4}

In this section, we will analyze the radial perturbation
about the scalarized BHs obtained in Sec. \ref{sec3}. 

\subsection{Master equation}
\label{sec41}

We start from the general time-dependent spherically symmetric 
ansatz of the spacetime and the scalar field:
\begin{align}
ds^2
&=
-\tilde A(t,r)dt^2
+\frac{dr^2}{\tilde B(t,r)}
+r^2 (d\theta^2+\sin^2\theta d\varphi^2),
\\
\phi&=\phi(t,r).
\end{align} 
We assume that the deviation from a static and spherically symmetric BH solution is small,
\begin{align}
\label{perturb}
{\tilde A}=A(r)+\varepsilon a(t,r),
\qquad
{\tilde B}=B(r)+\varepsilon b(t,r),
\qquad
{\phi}=\psi(r)+\varepsilon \Phi(t,r),
\end{align}
where $(A,B,\psi)$ represents a background BH solution
and $(a,b,\Phi)$ does the perturbation about it.
Expanding up to the first order of $\varepsilon (\ll 1)$,
the perturbed $(t,t)-$, $(r,r)-$, $(t,r)-$, and angular components of 
the gravitational equation of motion Eq.~\eqref{cov_grav_eq} 
and the perturbed scalar field equation of motion Eq.~\eqref{cov_sca_eq}
are, respectively, obtained as 
\begin{align}
& 
\label{pert1}
 \alpha_1 \Phi''
+ \alpha_2\Phi'
+\alpha_3\Phi
+\alpha_4 b
+\alpha_5 b'
=0 ,
\\
\label{pert2}
&
 \beta_1 \ddot{\Phi}
+\beta_2 \Phi'
+\beta_3 \Phi
+\beta_4 a
+\beta_5 a'
+\beta_6 b
=0,
\\
\label{pert3}
&
 \gamma_1 \dot{b}
+\gamma_2 \dot{\Phi}'
+\gamma_3 \dot{\Phi}
=0,
\\
\label{pert4}
&
 c_1 \ddot{\Phi}
+c_2 \Phi''
+c_3 \Phi'
+c_4 \Phi
+c_5 a''
+c_6 a'
+c_7 a
+c_8 \ddot{b}
+c_9 b'
+c_{10}b
=0,
\\
\label{pert5}
&
 d_1 \ddot{\Phi}
+d_2 \Phi''
+d_3 \Phi'
+d_4 \Phi
+d_5 a''
+d_6 a'
+d_7 a
+d_8 \ddot{b}
+d_9 b'
+d_{10}b
=0,
\end{align}
where 
a ``dot''  denotes the derivative with respect to the time $t$, 
and 
$\alpha_i$ ($i=1, 2, \cdots,5$),
$\beta_i$ ($i=1, 2, \cdots,6$),
$\gamma_i$ ($i=1,2,3$),
$c_i$ ($i=1, 2, \cdots,10$),
and
$d_i$ ($i=1, 2, \cdots,10$)
are the functions of $r$,
which are given in Appendix \ref{appa}.
Integrating \eqref{pert3} with respect to $t$,
\begin{align}
\label{b_eq}
  b
=
-\frac{1}{\gamma_1}
\left(\gamma_2 \Phi'
+\gamma_3 \Phi
\right),
\end{align}
where the integration constant is set to zero by a redefinition of the background solution.
Plugging Eq.~\eqref{b_eq} into Eq.~\eqref{pert2},
$a'$ can be expressed in terms of $a$, $\Phi$, $\Phi'$, and $\ddot{\Phi}$:
\begin{align}
\label{a_eq}
a'=
-\frac{\beta_1}{\beta_5} \ddot{\Phi}
+\frac{-\beta_2\gamma_1+\beta_6\gamma_2}{\beta_5\gamma_1}\Phi'
+\frac{-\beta_3\gamma_1+\beta_6\gamma_3}{\beta_5\gamma_1}\Phi
-\frac{\beta_4}{\beta_5}a.
\end{align}
Multiplying Eqs.~\eqref{pert4} and \eqref{pert5}
by $d_5$ and $c_5$,
respectively,
and taking their difference,
the term proportional to $a''$ is eliminated.
Substituting Eqs.~\eqref{b_eq} and \eqref{a_eq} into it
and eliminating $a'$ and $b$ (and its derivatives),
we finally arrive in the master equation for the radial perturbation: 
\begin{align}
\label{pert6}
-\rho_1 \ddot{\Phi}
+\rho_2 \Phi''
+\rho_3 \Phi'
+\rho_4 \Phi
=0,
\end{align}
where $\rho_i$ ($i=1, 2, 3, 4$) are given in Appendix \ref{appa}.
Assuming that $\Phi=e^{-i\omega t} \Phi_\omega (r)$ with $\Phi_\omega= C(r) \Psi_\omega(r)$
where the function $C(r)$ satisfies
\begin{align}
\frac{C'}{C}
=\frac{1}{4} (\ln (AB))'-\frac{\rho_3}{2\rho_2},
\end{align}
each mode satisfies the eigen equation
\begin{align}
\label{eigeneq_full}
\left[
-\frac{d^2}{dr_\ast^2}
+U_{\rm eff}(r)
\right]
\Psi_\omega(r)
=
\frac{\rho_1}{\rho_2}(AB)\omega^2 \Psi_\omega (r),
\end{align}
where we have introduced the tortoise radial coordinate $dr_\ast:= dr/\sqrt{AB}$,
and the effective potential
\begin{align}
\label{effpot_full}
U_{\rm eff}(r)
:=-AB 
\left\{
  \frac{1}{4}\left(\ln (AB)\right)''
+\frac{1}{16} \left[ \left(\ln (AB)\right)'\right]^2
-\frac{1}{2}
\left(\frac{\rho_3}{\rho_2}\right)'
-\frac{\rho_3^2}{4\rho_2^2}
+\frac{\rho_4}{\rho_2}
\right\}.
\end{align}
A scalarized BH solution is unstable,
if Eq.~\eqref{eigeneq_full} with Eq.~\eqref{effpot_full} admit pure imaginary modes $\omega^2<0$.
On the other hand, 
the absence of pure imaginary modes 
can be ensured,
if the effective potential $U_{\rm eff} (r)$
is non-negative everywhere outside the horizon $r\geq r_h$,
$U_{\rm eff}(r)\geq 0$. 
In Ref.~\cite{Blazquez-Salcedo:2018jnn},
the radial perturbation of the scalarized BHs in the scalar-tensor theory
with the quadratic order coupling to the GB term \eqref{quadratic}
was analyzed,
and the existence of pure imaginary modes was clarified,
indicating that the scalarized BH solutions for these coupling functions are unstable.
In the next subsection,
we will consider the case of the more general coupling \eqref{general}.

\subsection{Effective potential and stability}
\label{sec42}

We then apply the formulation in Sec. \ref{sec41}
for analyzing the stability of the scalarized BH solutions obtained in Sec. \ref{sec3}
against the radial perturbation.

\begin{figure}[h]
\unitlength=1.1mm
\begin{center}
  \includegraphics[height=7.5cm,angle=0]{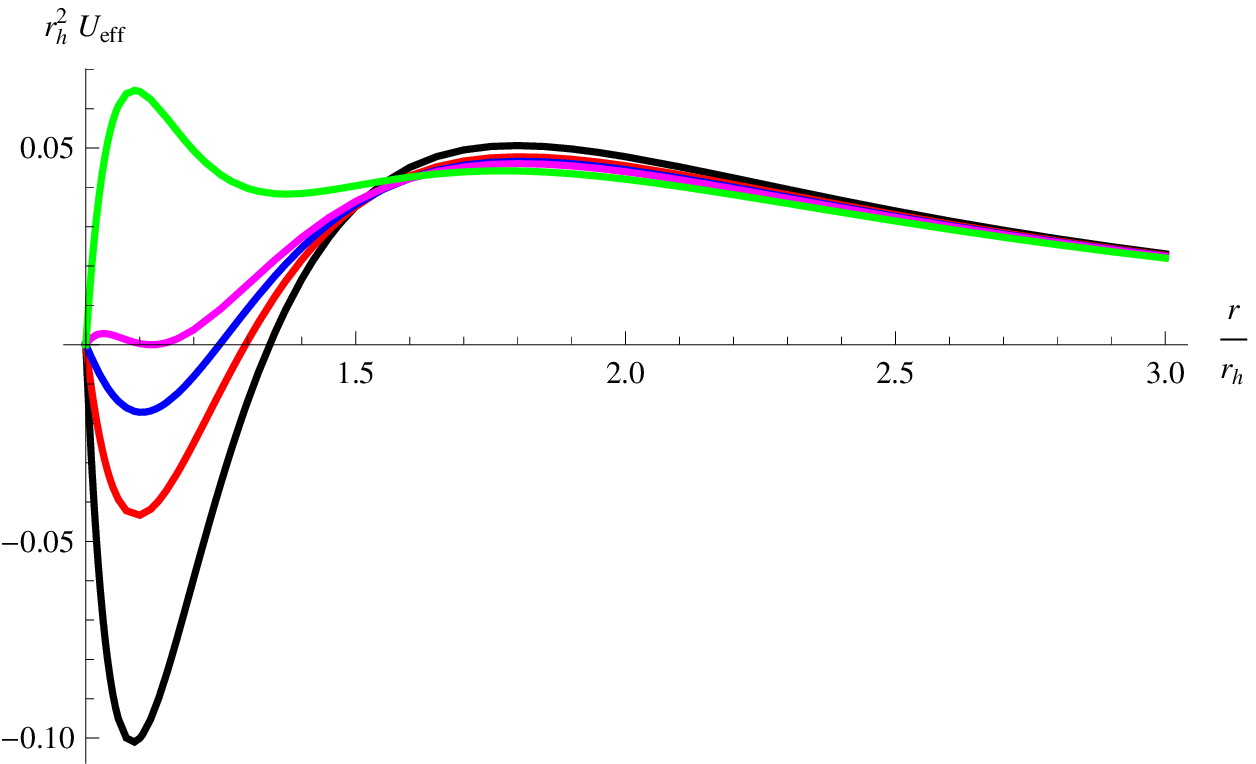}
  \includegraphics[height=5.0cm,angle=0]{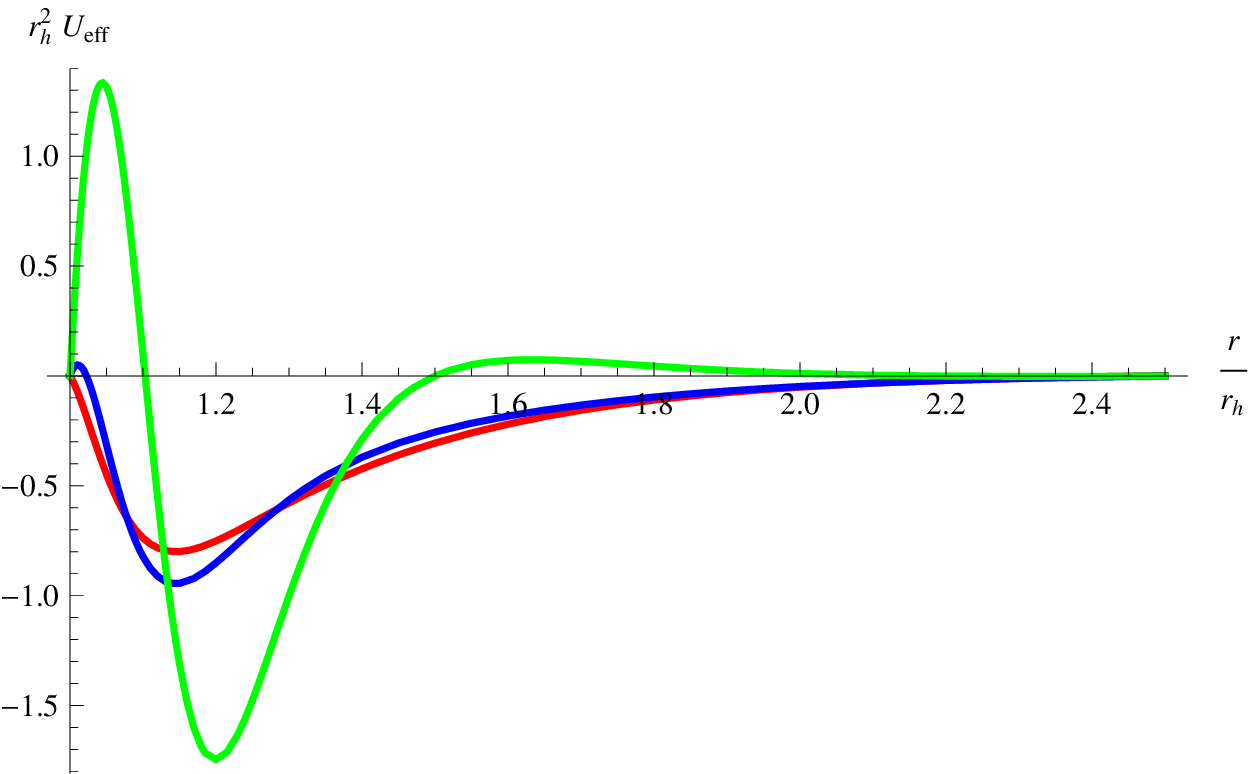}
  \includegraphics[height=5.0cm,angle=0]{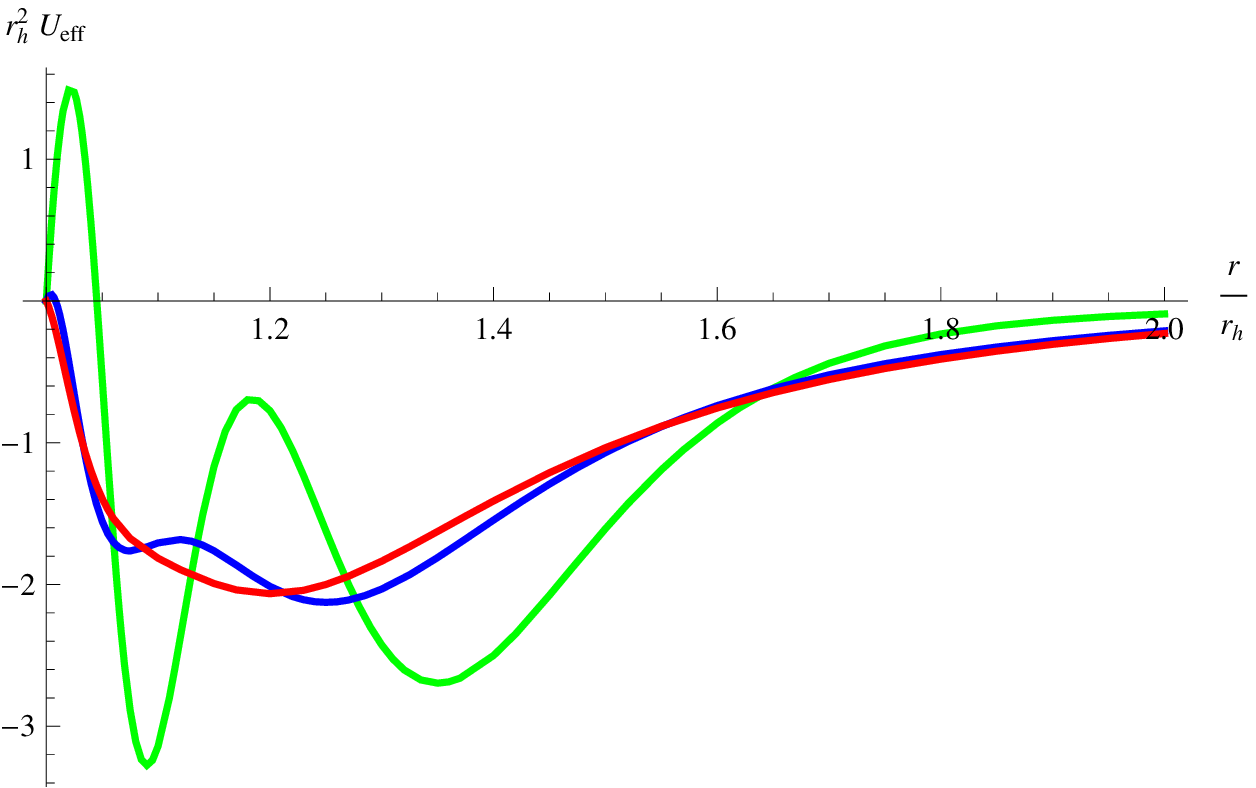}
\caption{
In the top panel,
the effective potential $U_{\rm eff} (r)$ multiplied by $r_h^2$
is shown as the function of $r/r_h$
for the scalarized BH solutions with zero nodes of the scalar field for $\psi_0=0.005$.
The black, red, blue, magenta, and green curves
correspond to 
the cases of the scalarized BHs 
for 
$(\eta/r_h^2,\alpha)=(0.558,10000)$, $(0.725,0)$, $(0.803,-3000)$, $(0.8505, -4619)$,
$(1.03,-9000)$,
respectively.
Similarly,
the bottom-left and bottom-right panels
describe
the effective potential $U_{\rm eff} (r)$ multiplied by $r_h^2$
as the function of the mass $r/r_h$
for the scalarized BH solutions with one and two nodes of the scalar field for $\psi_0=0.005$,
respectively.
In the bottom-left and bottom-right panels,
the red, blue, and green curves
correspond to the cases of 
for the scalarized BH solutions with one node
for $(\eta/r_h^2,\alpha)= (5.27,-4000)$, $(6.12,-10000)$, $(11.0,-19000)$,
and 
for the scalarized BH solutions with two nodes
for $(\eta/r_h^2,\alpha)= (13.4,-4000)$, $(14.9,-10000)$, $(22.9,-19000)$,
respectively.
}
  \label{figpotential}
\end{center}
\end{figure}

In the top panel of Fig.~\ref{figpotential},
the effective potential for the radial perturbation $U_{\rm eff}(r)$, Eq.~\eqref{effpot_full}, 
multiplied by $r_h^2$, is shown as the function of $r/r_h$
for the scalarized BH solutions with zero nodes of the scalar field  for $\psi_0=0.005$.
The black, red, blue, magenta, and green curves
correspond to 
the cases of the scalarized BH solutions 
for 
$(\eta/r_h^2,\alpha)=(0.558,10000)$, $(0.725,0)$, $(0.803,-3000)$, $(0.8505, -4619)$,
$(1.03,-9000)$, respectively.
In the case of $\alpha=0$,
$U_{\rm eff}(r)$ contains a negative region in the vicinity of the horizon $r=r_h$,
and the analysis in Ref. \cite{Blazquez-Salcedo:2018jnn} suggests
that it gives rise to a pure imaginary mode.
In the case of $\alpha>0$,
as shown in the black curve in the top panel of Fig.~\ref{figpotential},
the depth of the negative region of $U_{\rm eff}(r)$ 
becomes larger than in the case of $\alpha=0$,
and
we speculate
that the scalarized BH solutions obtained for $\alpha>0$ 
are also unstable against the radial perturbation
of those in the case of $\alpha=0$.

On the other hand, 
in the case of $\alpha<0$,
the minimum of $U_{\rm eff} (r)$ 
in the vicinity of the horizon $r=r_h$ 
increases
and
for a sufficiently large negative value of $\alpha$
no negative region appears,
as in the case of the green curve 
in the top panel of Fig.~\ref{figpotential}.
Thus,
only for a large negative quartic order coupling constant,
the scalarized BHs with zero nodes of the scalar field
are expected to become stable against the radial perturbation.
As shown by the magenta curve in the top panel of Fig.~\ref{figpotential},
the minimum of $U_{\rm eff}(r)$
becomes zero 
for $\alpha\approx -4619$ in the case of $\psi_0=0005$.
Hence, 
there is no negative region in $U_{\rm eff}(r)$
and 
the scalarized BH solutions with zero nodes of the scalar field
are expected to be stable for $-20000<\alpha\lesssim -4619$.
Moreover,
for $\psi_0=0.005$ and $\alpha\approx -4619$,
we find that $\alpha\psi_0^2\approx 0.1155$,
which almost agreed with the upper bound of Eq.~\eqref{alpha_bound}
in our test field analysis in Sec. \ref{sec13}.
Thus, 
at least for a small value of $\psi_0$,
the scalarized BH solutions with zero nodes of the scalar field 
which are stable against the radial perturbation
exist almost for the range given by Eq. \eqref{alpha_bound}.

In the bottom-left and bottom-right panels
of Fig.~\ref{figpotential},
the red, blue, and green curves
correspond to $U_{\rm eff} (r)$
for $(\eta/r_h^2,\alpha)= (5.27,-4000)$, $(6.12,-10000)$, $(11.0,-19000)$
for the scalarized BH solutions with one node,
and 
for $(\eta/r_h^2,\alpha)= (13.4,-4000)$, $(14.9,-10000)$, $(22.9,-19000)$
for the scalarized BH solutions with two nodes,
respectively.
We find that in the cases of the scalarized BHs 
with one and two nodes of the scalar field,
even for a negative value of $\alpha$,
the effective potential $U_{\rm eff}(r)$
always contains a negative region.
Thus,
we conclude 
that 
in the case of the coupling to the GB term \eqref{general}
the scalarized BH solutions with one and two nodes
are always unstable
against the radial perturbation.
We note that for all the cases shown in Fig.~\ref{figpotential},
$\rho_1>0$ and $\rho_2>0$ in Eq.~\eqref{pert6},
and, hence, there is no violation of the hyperbolicity of the radial perturbation,
at least for the given choice of the scalarized BH solutions.

The above features are quite similar to our perspectives obtained 
from the test field analysis in Sec. \ref{sec13}.
As we argued in Sec. \ref{sec32},
we speculate that 
this similarity 
is due to the presence of the bound \eqref{exist2},
namely, 
the amplitude of the scalar field cannot be so large
that the backreaction on the metric
significantly modifies features of the scalarized BHs.
As expected,
for the Schwarzschild solution with the constant scalar field $\psi(r)=\pm \sqrt{-1/(2\alpha)}$ for $\alpha<0$,
the effective potential $U_{\rm eff} (r)$ agrees with 
Eq.~\eqref{effpot_grsol} obtained in the test field analysis,
which contains no negative region.

Thus, 
we conclude that
among the scalarized BH solutions
in the scalar-tensor theory \eqref{general}
only the scalarized BH solutions with zero nodes 
for a sufficiently large negative value of $\alpha<0$
can be stable against the radial perturbation.  
We have to emphasize, however, that 
the radial stability does not ensure the stability of BHs against the other perturbations.
They should be analyzed separately in the future work.

\section{Conclusions}
\label{sec5}

We have investigated the scalarized black hole (BH) solutions
in the scalar-tensor theory 
with the coupling of the scalar field to the Gauss-Bonnet (GB) term \eqref{esgb}.
In Sec. \ref{sec1},
after reviewing how the spontaneous scalarization may take place around a vacuum BH 
due to a tachyonic instability triggered by the coupling to the GB term
which is analogous to that inside relativistic stars triggered by the coupling to the matter field,
we have demonstrated such a possibility by a test field analysis.
We have obtained the nontrivial solution of a static test scalar field 
which is coupled to the curvature of the background Schwarzschild spacetime.
We have shown
that around a Schwarzschild BH
there are the nontrivial static scalar field solutions 
with zero, one, and two nodes (and there should be those with more nodes
which were not investigated in this paper)
for the pure quadratic order coupling \eqref{quadratic}
and the general coupling \eqref{general}.
We have also analyzed the radial scalar field perturbation of these nontrivial scalar field solutions,
and 
confirmed the existence of a negative region of the effective potential of the scalar field perturbation.
Only the exceptional case was the scalar field solution with zero nodes
for a sufficiently large negative value of the quartic order coupling satisfying Eq. \eqref{alpha_bound},
where there is no negative region in the effective potential and hence no pure imaginary mode.

In Sec.~\ref{sec2},
we have explained our strategy for constructing
the static and spherically symmetric scalarized BH solutions 
in the scalar-tensor theory with an arbitrary coupling to the GB term.
By arranging the components of the gravitational equations of motion \eqref{eq1} and \eqref{eq2}
and the scalar field equation of motion \eqref{eq3},
we have obtained a set of the equations \eqref{eq_A} and \eqref{eq_psi} 
to determine 
the time component of the metric $A(r)$ and the scalar field $\psi(r)$.
With use of the solution for $A(r)$ and $\psi(r)$, 
the radial component of the metric $B(r)$ was then determined via Eq.~\eqref{b}.
We have focused on the scalarized BH solutions 
whose asymptotic value of the scalar field at the infinity is zero,
which may be the endpoint of the tachyonic instability
of the Schwarzschild BH with the constant scalar field $\psi(r)=0$.

In Sec. \ref{sec3},
we have numerically constructed the 
scalarized BH solutions 
in the scalar-tensor theories 
with
the pure quadratic order coupling \eqref{quadratic}
and the more general coupling \eqref{general} to the GB term.
The case of the pure quadratic order coupling
has been already analyzed in Ref. \cite{Silva:2017uqg},
and 
the results were recovered in our analysis.
In the case of the more general coupling \eqref{general},
we numerically confirmed the existence of 
the scalarized BH solutions with zero, one, two nodes
(and there should be the solutions with more nodes),
as well as the Schwarzschild BH solutions with the constant scalar field 
$\psi(r)=0$ and $\psi(r)=\pm \sqrt{-1/(2\alpha)}$,
where the latter exists only for $\alpha<0$.
In the limit of $\alpha\gg 1$,
if the amplitude of the scalar field at the horizon $\psi_0$ is below the certain critical value
given by Eq.~\eqref{psi_bound},
the scalarized BH solutions approach those in the case of the pure quartic order coupling \eqref{quartic}.
If $\psi_0$ is larger than the value \eqref{psi_bound},
there is the maximal value of $\alpha$
above which there is no scalarized BH solution
and hence no limit to the pure quartic order coupling model.
For  $\alpha<0$,
there were also the scalarized BH solutions,
but, 
in order to satisfy the boundary condition,
$\alpha$ had to satisfy the bound \eqref{general2} for a given $\psi_0$.

In Sec. \ref{sec4},
we have investigated the radial perturbation of the scalarized BHs obtained in Sec. \ref{sec3}.
The combination of the nontrivial components of the gravitational equations of motion
and the scalar field equation of motion
resulted in the single master equation \eqref{pert6}.
We have confirmed
that 
in the case of the pure quadratic order coupling
the effective potential for the radial perturbation \eqref{effpot_full}
has a negative region in the vicinity of the horizon $r=r_h$
for all the scalarized BHs with zero, one, and two nodes,
indicating the existence of pure imaginary modes.
We have shown
that in the case of the general coupling \eqref{general}
with the negative quartic order coupling constant $\alpha<0$,
only for the scalarized BH solutions with zero nodes of the scalar field 
the effective potential becomes non-negative in the vicinity of the horizon $r=r_h$,
implying the absence of pure imaginary modes.
This was consistent with the test field analysis performed in Sec. \ref{sec13}.
We also expected that 
the region where the scalarized BH solutions are
stable against the radial perturbation
coincides with 
that obtained in the test field analysis, Eq. \eqref{alpha_bound},
at least for a smaller value of $\psi_0$.
On the other hand, 
the scalarized BH solutions with more than one node of the scalar field
were always expected to be unstable against the radial perturbation.

There will be a number of subjects related to our work.
First,
for the scalarized BH solutions 
which were expected to be stable against the radial perturbation,
in order to clarify the complete stability,
it will be necessary
to analyze the other type of perturbations.
On the other hand,
in the case that the scalarized BH solutions are unstable against the radial perturbation,
the scalarized BHs would not be the fate of gravitational collapse.
In such a case, 
it will be very interesting
to develop a numerical scheme 
to follow gravitational collapse of the scalar field.
Finally,
it will also be interesting 
to consider the scalarization of a vacuum BH solution 
in the presence of the coupling of the scalar field to other invariants,
such as
the Pontryagin density ${}^\ast R^{\mu\nu\alpha\beta} R_{\mu\nu\alpha\beta}$
and the Maxwell term $F_{\mu\nu}F^{\mu\nu}$ \cite{Herdeiro:2018wub}
(See also for the earlier works on the coupling to the Born-Infeld term \cite{Stefanov:2007eq,Doneva:2010ke}).
We will come back to these issues in future publications.

\acknowledgments{
We thank Thomas Sotiriou for helpful discussions.
M.~M. was supported by FCT-Portugal through Grant No.\ SFRH/BPD/88299/2012. 
T.~I. acknowledges financial support provided under the European Union's H2020 ERC Consolidator Grant ``Matter and strong- field gravity: New frontiers in Einstein's theory'' grant agreement no. MaGRaTh-646597, and under the H2020-MSCA-RISE-2015 Grant No. StronGrHEP-690904.  
}

\appendix

\section{Coefficients}
\label{appa}

In this appendix,
we show the coefficients in the equations 
for the perturbative analysis shown in Sec. \ref{sec4}.

The coefficients in Eq.~\eqref{pert1}
are given by 
\begin{align}
\alpha_1
&=
4(B-1)B \xi^{(1)}(\psi),
\\
\alpha_2
&=
\frac{1}{2}
\left(
4(3B-1)B' \xi^{(1)}(\psi)
+B \psi'
(-r^2+16 (B-1) \xi^{(2)}(\psi))
\right),
\\
\alpha_3
&=
2
\left[
 (3B-1)B' \psi' \xi^{(2)} (\psi)
+2 (B-1)B
\left(
  \xi^{(2)}(\psi)\psi''
+\psi'^2\xi^{(3)}(\psi)
\right)
\right],
\\
\alpha_4
&=
\frac{1}{4}
\left[
  24B' \xi^{(1)}(\psi)\psi'
-\psi'^2 
\left(
r^2+
16(1-2B)\xi^{(2)} (\psi)
\right)
+4
\left(
-1+4 (2B-1)
\xi^{(1)}(\psi)\psi''
\right)
\right],
\\
\alpha_5
&=
-
\left(
r+2(1-3B)\xi^{(1)} (\psi)\psi'
\right).
\end{align}
The coefficients in Eq.~\eqref{pert2}
are given by 
\begin{align}
\beta_1
&=-\frac{4(B-1)\xi^{(1)}(\psi) }{A},
\\
\beta_2
&=
\frac{B}{2 A}
\left[
4(3B-1)A' \xi^{(1)}(\psi)
+r^2 A\psi'
\right],
\\
\beta_3
&=\frac{2 B(3B-1)A'\psi'\xi^{(2)}(\psi)}{A},
\\
\beta_4
&=
\frac{BA'}{A^2}
\left(
r+2(1-3B)\xi^{(1)} (\psi)\psi'
\right),
\\
\beta_5
&=
-\frac{B}{A}
\left(
r+2(1-3B)\xi^{(1)} (\psi)\psi'
\right),
\\
\beta_6
&=
\frac{1}{4 A}
\left[
-4A' 
\left(
r+2(1-6B)\xi^{(1)}(\psi)\psi'
\right)
+A (-4+r^2\psi'^2)
\right].
\end{align}
The coefficients in Eq.~\eqref{pert3}
are given by 
\begin{align}
\gamma_1
&=\frac{r + 2(1-3B)\xi^{(1)}(\psi)\psi'}
        {B},
\\
\gamma_2
&=-4(B-1)\xi^{(1)} (\psi),
\\
\gamma_3
&=
\frac{1}{2A}
\left[
4(B-1)A' \xi^{(1)}(\psi)
+A\psi'
\left(
r^2 -8(B-1)\xi^{(2)}(\psi)
\right)
\right].
\end{align}
The coefficients in Eq.~\eqref{pert4}
are given by 
\begin{align}
c_1
&=
-\frac{2B' \xi^{(1)} (\psi)}{A},
\\
c_2
&=
\frac{2B^2 A' \xi^{(1)} (\psi)}{A},
\\
c_3
&=\frac{B}{2A^2}
\left[
 A (6A'B'\xi^{(1)} (\psi)-rA \psi')
+B
(-2A'^2 \xi^{(1)}(\psi)
+4A\xi^{(1)}(\psi) A''
+8AA' \psi' \xi^{(2)}(\psi))
\right],
\\
c_4
&=
\frac{B}{A^2}
\left[
3AA' B'\psi'\xi^{(2)}(\psi)
+B
\left(
-A'^2 \psi'\xi^{(2)}(\psi)
+2A\psi' A''\xi^{(2)}(\psi)
+2AA' 
(\xi^{(2)} (\psi)\psi''+ \psi'^2 \xi^{(3)}(\psi))
\right)
\right],
\\
c_5
&=
-\frac{B (r-4B\xi^{(1)}(\psi)\psi')}{2A},
\\
c_6
&=
\frac{1}{4A^2}
\left[
2BA'(r-4B\xi^{(1)} (\psi)\psi')
\right.
\nonumber\\
&
\left.
+A
\left(
-rB'+2B (-1+6B'\xi^{(1)}(\psi)\psi')
+8B^2 (\psi'^2\xi^{(2)}(\psi)+\xi^{(1)}(\psi)\psi'')
\right)
\right],
\\
c_7
&=
\frac{1}{4A^3}
\left[
2BA'^2
(-r+4B\xi^{(1)} (\psi)\psi')
\right.
\nonumber\\
&\left.
+A
\left(
 rA' B'
+2B
\left(A'(1-6B'\xi^{(1)}(\psi) \psi')
+rA''
\right)
-8B^2
 \left(
  A' \psi'^2\xi^{(2)}(\psi)
+\xi^{(1)}(\psi) 
  (\psi' A''+A'\psi'')
 \right)
\right)
\right],
\\
c_8
&=
-\frac{r-4B\xi^{(1)}(\psi)\psi' }
        {2AB},
\\
c_9
&=
-\frac{1}{4A}
\left(
2A +A' (r-12B\xi^{(1)} (\psi)\psi')
\right),
\\
c_{10}
&=
-\frac{1}{4A^2}
\left[
 rA^2 \psi'^2
-A'^2 (r-8B\xi^{(1)} (\psi)\psi')
\right.
\nonumber\\
&\left.
-2A
\left(
-(r-8B\xi^{(1)}(\psi)\psi')A''
+A'
\left(
-1
+6B'\xi^{(1)}(\psi)\psi'
+8B(\xi^{(2)} (\psi)\psi'^2+\xi^{(1)} (\psi)\psi'')
\right)
\right)
\right].
\end{align}
The coefficients in Eq.~\eqref{pert5}
are given by 
\begin{align}
d_1
&=-\frac{1}{A},
\\
d_2
&= B,
\\
d_3
&=\frac{1}{2}
\left(
B\left(\frac{4}{r}+\frac{A'}{A}\right)
+B'
\right),
\\
d_4
&=
-\frac{2\xi^{(2)} (\psi)}{r^2A^2}
\left[
  A A' B'
+B^2 (A'^2-2A A'')
-B 
(A'^2+3AA' B'-2A A'')
\right],
\\
d_5
&=
\frac{4(B-1)B\xi^{(1)}(\psi)}{r^2A},
\\
d_6
&=
\frac{1}{2r^2 A^2}
\left[
-8B^2 A' \xi^{(1)} (\psi)
-4A B' \xi^{(1)} (\psi)
+B
\left(
 8A' \xi^{(1)} (\psi)
+A (12 B' \xi^{(1)}(\psi)+r^2\psi')
\right)
\right],
\\
d_7
&=\frac{1}{2r^2A^3}
\left[
  4AA' B'\xi^{(1)}(\psi)
+ 8 B^2\xi^{(1)}(\psi)
  (A'^2-AA'')
\right.
\nonumber\\
&\left.
-B
\left(
  8A'^2 \xi^{(1)}(\psi)
+AA' (12B'\xi^{(1)} (\psi)+r^2\psi')
-8A\xi^{(1)}(\psi) A''
\right)
\right],
\\
d_8
&=
\frac{4(B-1)\xi^{(1)}(\psi)}
       {r^2 A B},
\\
d_9
&=\frac{1}{2r^2A}
\left(
 4(3B-1)A'\xi^{(1)}(\psi)
+r^2A\psi'
\right),
\\
d_{10}
&=
\frac{1}{2r^2A^2}
\left[
 4 (1-2B)A'^2 \xi^{(1)}(\psi)
+AA' (12B'\xi^{(1)}(\psi)+r^2\psi')
+2A 
\left(
 4(2B-1)\xi^{(1)}(\psi)A''
+rA (2\psi'+r\psi'')
\right)
\right].
\end{align}
The coefficients in Eq.~\eqref{pert6} are given by 
\begin{align}
\rho_1
&=
-d_5
\left(
c_1
-\frac{c_6 \beta_1}{\beta_5}
-\frac{c_8\gamma_3}{\gamma_1}
\right)
+
c_5
\left(
d_1
-\frac{d_6 \beta_1}{\beta_5}
-\frac{d_8\gamma_3}{\gamma_1}
\right),
\\
 \rho_2
&=\frac{1}{\gamma_1} 
\left(
  d_5(c_2\gamma_1-c_9\gamma_2)
-c_5 (d_2\gamma_1-d_9 \gamma_2)
\right),
\\
\rho_3
&=\frac{1}{\beta_5\gamma_1^2}
\left[
  c_6 d_5 \gamma_1 (-\beta_2\gamma_1+\beta_6\gamma_2)
+d_5\beta_5
\left(
  c_3\gamma_1^2-c_{10}\gamma_1\gamma_2
-c_9(-\gamma_2\gamma_1'+\gamma_1(\gamma_3+\gamma_2'))
\right)
\right.
\nonumber\\
&
\left.
-c_5 d_6\gamma_1 (-\beta_2\gamma_1+\beta_6\gamma_2)
-c_5\beta_5
\left(
  d_3\gamma_1^2-d_{10}\gamma_1\gamma_2
-d_9(-\gamma_2\gamma_1'+\gamma_1(\gamma_3+\gamma_2'))
\right)
\right],
\\
\rho_4
&=
\frac{1}{\beta_5\gamma_1^2}
\left[
 c_6 d_5\gamma_1
 (-\beta_3\gamma_1+\beta_6\gamma_3)
+d_5 \beta_5
(c_4\gamma_1^2
-c_{10}\gamma_1\gamma_3
+c_9 \gamma_3\gamma_1'
-c_9 \gamma_1\gamma_3')
\right.
\nonumber\\
&\left.
-d_6 c_5\gamma_1
 (-\beta_3\gamma_1+\beta_6\gamma_3)
-c_5 \beta_5
(d_4\gamma_1^2
-d_{10}\gamma_1\gamma_3
+d_9 \gamma_3\gamma_1'
-d_9 \gamma_1\gamma_3')
\right].
\end{align}

\bibliography{ref-nohair}

\end{document}